\RequirePackage{fix-cm}
\documentclass[smallcondensed]{svjour3ujm}
\usepackage{graphicx}

\begin{document}
\raggedbottom
\title{Bohr, QBism, and Beyond}
\author{Ulrich J. Mohrhoff}
\institute{Ulrich J. Mohrhoff\\
Sri Aurobindo International Centre of Education\\Pondicherry 605002 India\\
\email{ujm@auromail.net}\\
}
\maketitle
\vskip-36pt
\begin{abstract}
QBism may be the most significant contribution to the search for meaning in quantum mechanics since Bohr, even as Bohr's philosophy remains the most significant revision of Kant's theory of science. There are two ironies here. Bohr failed to realize the full extent of the affinity of his way of thinking with Kant's, and QBists fail to realize the full extent of their agreement with Bohr. While Bohr's discovery of contextuality updates Kant's transcendental philosophy in a way that leaves the central elements of the latter intact, Kant's insight into the roles that our cognitive faculties play in constructing physical theories can considerably alleviate the difficulties that Bohr's writings present to his readers. And while throwing a QBist searchlight on Bohr's writings can further alleviate these difficulties (as well as reveal the presence in them of the salient elements of QBist thought), Bohr's writings can in turn provide answers to important questions that QBism leaves unanswered (and also allay some of QBism's excesses and possible inconsistencies). In the final sections I confront the two most impenetrable mysteries yet unearthed: making sense of quantum mechanics, and the dual mystery of making sense of (i) the existence of consciousness in a seemingly material world and (ii) the existence in consciousness of a seemingly material world. Here the relevant arguments are framed in the context of the philosophy of the Upanishads, according to which we (as Schr\"odinger put it) ``are all really only various aspects of the One.'' There is no world that exists out of relation to consciousness, but there are different poises of consciousness. In particular, there is a poise of consciousness peculiar to the human species at this point in time, and there are poises of consciousness that are yet to evolve (and that may be essential to averting the calamities towards which humanity appears to be heading).
\end{abstract}

\section{Introduction}\label{sec.intro}
Quantum mechanics has a well-known problem. It comes in two versions, a BIG one and a \textsc{small} one.\cite{Pitowsky2006} The former calls for an explanation of how measurement outcomes come about dynamically. It is a pseudo-problem if ever there was one. Pseudo-problems arise from false assumptions, in this case the belief that a quantum state is some kind of evolving physical state. The latter arises once it is acknowledged that the mathematical apparatus of quantum mechanics is a probability calculus, and that the events to which (and on the basis of which) it assigns probabilities are measurement outcomes. It calls for a demonstration of consistency between the representation of measurement outcomes by subspaces of a Hilbert space and the representation of outcome-indicating states or events by atomic subsets of a phase space. Nine decades after this problem was solved by Niels Bohr in or around 1929 (albeit in a way that nobody seems to have understood), and after half a century of futile attempts at solving it without taking account of the universal context of science, which is human experience,\cite{Schilling} there is light at the end of the tunnel. It is called QBism. Launched at the beginning of the 21st Century by {Carlton Caves}, {Chris Fuchs}, and {Ruediger Schack},\cite{CFS2002} QBism may be the most significant contribution to the search for meaning in quantum mechanics since Bohr, even as Bohr's philosophy of quantum mechanics remains the most significant revision of Kant's theory of science. 

To David Mermin,\cite{MerminQBnotCop} QBism is ``as big a break with 20th century ways of thinking about science as Cubism was with 19th century ways of thinking about art.'' The big break lies not in the emphasis that the mathematical apparatus of quantum mechanics is a probability calculus---that ought to surprise no one---but in this \emph{plus} a radically subjective Bayesian interpretation of probability \emph{plus} a radically subjective interpretation of the events to which (and on the basis of which) probabilities are assigned. What distinguishes the outcome-indicating properties of outcome-indicating devices from other physical properties is that they are \emph{perceived}. They are \emph{experiences}. Nothing but the incontestable definiteness and irreversibility of direct sensory experience can account for the definiteness of outcome-indicating properties and the irreversibility of measurements.

There are two ironies here. The first is that Bohr failed to realize the full extent of the affinity of his way of thinking with Kant's. The second is that QBists fail to realize the full extent of their agreement with Bohr. While Bohr's discovery of the contextuality of quantum phenomena updates Kant's transcendental philosophy in ways that leave the central elements of the latter intact, being acquainted with Kant's insight into the roles that our cognitive faculties of intuition (\emph{Anschauung}) and thought play in constructing physical theories can considerably alleviate the difficulties that Bohr's writings present to his readers. And while throwing a QBist searchlight on Bohr's writings can further alleviate these difficulties (as well as reveal the presence in them of the salient elements of QBist thought), Bohr's writings can in turn provide answers to important questions that QBism leaves unanswered (and also allay some of QBism's disconcerting extravagances and possible inconsistencies).

My first order of business, carried out in Sect.~\ref{sec.3realisms}, is to set off empirical realism---the kind of realism that was inaugurated by Immanuel Kant and defended (among others) by Hilary Putnam and Bernard d'Espagnat%
\footnote{Putnam assumed the existence of a mind-independent real world but  insisted that it does not dictate its own descriptions to us: ``talk of ordinary empirical objects is not talk of things-in-themselves but only talk of things-for-us''\cite{Putnam81}; ``we don't know what we are talking about when we talk about `things in themselves'.''\cite{Putnam87} D'Espagnat\cite{d'Espagnat_VR} stressed the necessity of distinguishing between an empirically inaccessible \textit{veiled} reality and an intersubjectively constructed \textit{empirical} reality.}---%
against the two kinds of realism that preceded it: direct or na\"{\i}ve realism and indirect or representational realism. Section~\ref{sec.Kant} presents in outline Kant's transcendental philosophy, and  Sect.~\ref{sec.KantQT} backs up the claim that Bohr's philosophy of quantum mechanics agrees in all essential respects with Kant's theory of science. 

Bohr's unique understanding of quantum mechanics is the focus of the next four sections, beginning in Section \ref{sec.Bohr} with a general outline of his views. It is important to distinguish Bohr's views from (all variants of) the Copenhagen interpretation. This interpretation only emerged in the mid-1950's, in response to David Bohm's hidden-variables theory and the Marxist critique of Bohr's alleged idealism, which had inspired Bohm.\cite{Chevalley99} The term ``Copenhagen interpretation'' first appeared in print in Heisenberg's version of 1955.\cite{Heisenberg1955}

Section~\ref{sec.QsysQpha} highlights a key implication of Bohr's thought, which is that the dichotomy between quantum systems and attributes created for them by measurements is unwarranted: experimental conditions are constitutive not only of the attributes of quantum systems but also of the systems themselves. Section~\ref{sec.OLCC} aims to clarify the respective roles that ordinary language and classical concepts play in Bohr's thought, and Sect.~\ref{sec.amplification} addresses a major stumbling block that Bohr's writings present to the reader, i.e., his several invocations of ``irreversible amplification effects.''

The next four sections are centered around QBism. Section~\ref{sec.QBLEW} contrasts the role that language plays (or is claimed to play) in QBism with the role it plays in Bohr's thinking. Section~\ref{sec.boundary} addresses certain flaws in QBism that quite unnecessarily distract from its core message. Section~\ref{sec.OnotR} brings up QBists' (widely shared) misappreciation of Bohrian thought, and Sect.~\ref{sec.w-i-i} raises the question of whether QBism countenances a reality beyond ``the common external world we have all negotiated with each other''.\cite{MerminQBnotCop} (The jury appears to be still out on this.)

In Sect.~\ref{sec.realcrit} I revisit a reality criterion I previously proposed for distinguishing between two kinds of observables: the contextual ones that have values only when measured, and the ones whose values exist independently of measurements (and thus are capable of indicating measurement outcomes). While superior to appeals to the size or weight of the measurement apparatus in justifying the irreversibility of measurements (as Bohr \emph{seems} to have done), this criterion cannot establish more than the empirical reality of an intersubjectively constructed world. It only permits us to treat the known world \emph{as if} it existed independently of the subjects knowing it.

The sections that follow present my attempts to confront the two most impenetrable mysteries we have yet unearthed: making sense of quantum mechanics, and the dual mystery of making sense of (i)~the existence \emph{of} consciousness \emph{in} a seemingly material world and (ii)~the existence \emph{in} consciousness \emph{of} a seemingly material world. To my mind, these mysteries are so intertwined that neither of them can be solved in isolation. 

Section~\ref{sec.identity} offers an explanation for ``the miraculous identity of particles of the same type,'' which according to Misner \emph{et al.}\cite{Misneretal1215} ``must be regarded, not as a triviality, but as a central mystery of physics.'' If correct, it not only implies the \emph{numerical} identity of particles of the same type but also makes it possible to argue that, at bottom, any object we observe \emph{here} with \emph{these} properties and any object we observe \emph{there} with \emph{those} properties are one and the same ``thing.'' It further suggests that quantum physics concerns not the world (as classical physics does) but \emph{how the world is manifested to us}. Here the relevant arguments are framed in the context of the philosophy of the Upanishads, according to which we (as Schr\"odinger phrased it) ``are all really only various aspects of the One'',\cite{SchrWhatIsReal} and which is outlined in Sect.~\ref{sec.poises}.

Among the kinds of cognition posited by the Upanishads, one deserves special attention, to wit, indirect knowledge, which is mediated by representations. This forms the subject of Sect.~\ref{sec.indirect}. In Sect.~\ref{sec.whyqm} the role that quantum mechanics plays in said context is examined, and an answer to the question why the fundamental theoretical framework of contemporary physics is probability calculus is proposed. In Sect.~\ref{sec.evolution} an Upanishadic theory of evolution is outlined, and Sect.~\ref{sec.future} ventures to set forth a possible future.

\section{Three kinds of realism: direct, representational, and empirical}\label{sec.3realisms}
In an essay written during the last year of his life,\cite{SchrWhatIsReal} Erwin Schr\"odinger expressed his astonishment at the fact that despite ``the absolute hermetic separation of my sphere of consciousness'' from everyone else's, there is ``a far-reaching structural similarity between certain parts of our experiences, the parts which we call external; it can be expressed in the brief statement that we all live in the same world.'' This similarity, Schr\"odinger avowed, ``is not rationally comprehensible. In order to grasp it we are reduced to two irrational, mystical hypotheses,'' one of which is ``the so-called {hypothesis of the real external world}''.%
\footnote{See Sect.~\ref{sec.whyqm} for the second of the two hypotheses.}
Schr\"odinger left no room for uncertainty about what he thought of this hypothesis: to invoke ``the existence of a real world of bodies which are the causes of sense-impressions and produce roughly the same impression on everybody \dots\ is not to give an explanation at all; it is simply to state the matter in different words. In fact, it means laying a completely useless burden on the understanding.'' For while we can compare the ``external'' contents of our respective spheres of consciousness through communication, we have no access to this real world of bodies and no way of knowing how it relates to those parts of our experiences about which we agree. 

Before Descartes, to \emph{be} was either to be a substance or to be a property of a substance. With Descartes, the human conscious subject assumed the role of a substance: to \emph{be} meant either to be a subject or to exist as a representation for a subject. Thus was born the representative theory of perception, and along with it the aforesaid completely useless burden on the understanding.

\begin{figure}[t]\begin{center}
\includegraphics[width=4.5in]{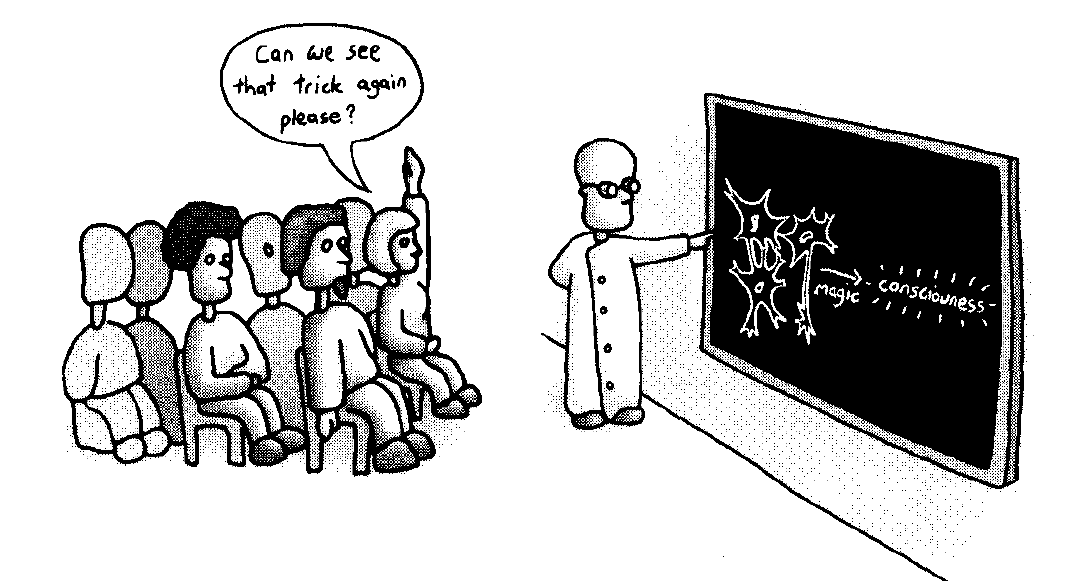}
\caption{A neuroscientist explaining the explanatory gap. Drawing by Jolyon Troscianko (jolyon.co.uk). Reproduced with permission.}\label{fig.trick}
\end{center}\end{figure}
Most current scientific accounts of perception still labor under this burden. They begin by assuming the existence of a mind-independent external world, in which objects emit photons or sound-waves, which stimulate peripheral nerve endings (retinas or ear drums). The stimulated nerves then send signals to the brain, where neural processes miraculously give rise to perceptual experience (Fig.~\ref{fig.trick}). Neither do we have the slightest idea of how this ``explanatory gap''\cite{Levine2001} is bridged, nor do we have the slightest idea of how we could have knowledge of what goes on in this mind-independent external world. While the aforesaid scientific accounts begin by invoking events in such a world, they lead to the conclusion that we have access only to perceptual experience.%
\footnote{This was already obvious to the Greek philosopher-poet Xenophanes, who some twenty-five centuries ago pointed out that even if our minds represented the world exactly as it was, we could never know that it did.}

In the eyes of John Searle,\cite[p.~23]{Searle2004}  the move from the na\"{\i}ve view that ``we really perceive real objects'' to the view that we only perceive sense-impressions, was ``the greatest single disaster in the history of philosophy over the past four centuries.'' In an attempt to defend the earlier na\"{\i}ve or direct realism against this indirect or representational realism, he invoked the fact that we are able to communicate with other human beings using publicly available meanings in a public language. For this to work, we have to assume the existence of common, publicly available objects of reference\cite[p.~276]{Searle2004}: 
\begin{quote}
So, for example, when I use the expression ``this table'' I have to assume that you understand the expression in the same way that I intend it. I have to assume we are both referring to the same table, and when you understand me in my utterance of ``this table'' you take it as referring to the same object you refer to in this context in your utterance of ``this table.''
\end{quote}
The implication then is that 
\begin{quote}
you and I share a perceptual access to one and the same object. And that is just another way of saying that I have to presuppose that you and I are both seeing or otherwise perceiving the same public object.  But that public availability of that public world is precisely the direct realism that I am here attempting to defend.
\end{quote}
Searle points out that his argument is transcendental in the sense of Immanuel Kant. A transcendental argument begins by assuming that a certain proposition \textbf{p} is true, and then shows that another proposition \textbf{q}, stating a precondition for the truth of \textbf{p}, must also be true. For Kant the relevant proposition \textbf{p} was the assumption that empirical knowledge was possible, and the corresponding proposition \textbf{q} was the conclusion that certain universal laws of nature must hold. In the argument presented by Searle, \textbf{p} is the assumption that we are able to communicate with each other in a public language, and \textbf{q} is the conclusion that there must be publicly available objects in a public world about which we can communicate in a public language. 

The realism that Searle's argument succeeds in defending is not the one it purports to defend. It is the empirical realism that was inaugurated by Kant\cite{KantCPR} and defended by Putnam\cite{Putnam81,Putnam87} and d'Espagnat\cite{d'Espagnat_VR} (among others). It is neither the na\"{\i}ve realism that reifies the perceived world nor a realism based on agreement between a mental construct or representation and a reality independent of us, but a realism based on agreement between our respective ``spheres of consciousness''---between what exists for me, in my experience, and what exists for you, in your experience. 

\section{Kant}\label{sec.Kant}
{\leftskip\parindent{[T]hose who really want to understand contemporary physics---i.e., not only to apply physics in practice but also to make it transparent---will find it useful, even indispensable at a certain stage, to think through Kant's theory of science.\par\hfill--- Carl Friedrich von Weizs\"acker}\cite{vW1980-328}\par}
\medskip\noindent
Transcendental philosophy---inaugurated by Kant and continued in the 20th century by Edmund Husserl\cite{HusserlEssential} and others---emerged as a critique of the representative theory. Here is how it was defined by Kant: ``I call all cognition transcendental that is occupied not so much with objects but rather with our mode of cognition of objects insofar as this is to be possible \emph{a priori}. A system of such concepts would be called transcendental philosophy.''[CPR\,149]%
\footnote{These references are to \emph{The Critique of Pure Reason}.\cite{KantCPR}} The concepts in question are synthetic rather than analytic. They are synthetic in that they enable us to ``work up the raw material of sensible impressions into a cognition of objects''[CPR\,136], and they are not analytic because they owe nothing to contingent experience. They owe their meanings to the logical structure of thought and the spatiotemporal structure of human sensory perception. 

The logical relation between a subject and a predicate makes it possible to think of a particular nexus of appearances as the properties of a \emph{substance}, connected to it as predicates are connected to a subject. It makes it possible for me to think of my perceptions as connected not in or by me, the subject, but in an external object. The logical relation between antecedent and consequent (if \dots\ then\dots) makes it possible to think of appearances at different times as events or properties connected by \emph{causality}. It makes it possible for me to think of successive perceptions as connected not merely in my experience but objectively, in an external world. And the category of community or reciprocity, which Kant associated with the disjunctive relation (either\dots\ or\dots), makes it possible to think of appearances in different locations as events or properties connected by a \emph{reciprocal action}. It makes it possible for me to think of simultaneous perceptions as objectively connected. (Kant thought that by establishing a reciprocal relation, we establish not merely an objective spatial relation but also an objective relation of simultaneity.)

But if I am to be able to think of my perceptions as a system of external objects, the connections must be lawful. If  appearances are to be perceptions of a particular kind of object (say, an elephant), they must be connected in an orderly way, according to a concept denoting a lawful concurrence of appearances. If appearances are to be perceptions of causally connected events, like (say) lightning and thunder, they must fall under a causal law, according to which one appearance necessitates the subsequent occurrence of another. (By establishing a causal relation falling under a causal law, we also establish an objective temporal relation.) And if appearances are to be reciprocally connected objects, like (say) the Earth and the Moon, they must affect each other according to a reciprocal law, such as Newton's law of gravity. It is through lawful connections in the ``manifold of appearances'' that we are able to think of appearances as perceptions of a self-existent system of objects. 

Kant's inquiry into the preconditions of empirical science was an inquiry into the preconditions of the possibility of organizing sense-impressions into objects---things that the subjects of these impressions could treat as if they existed independently of subjects and their impressions. The crucial premise of this inquiry was that ``space and time are only forms of sensible intuition, and therefore only conditions of the existence of the things as appearances.''[CPR\,115] They are not conditions of the existence of things in themselves, things that exist independently of subjects and their impressions. Combined with the fact that all physical concepts have visualizable content, and thus owe their meanings to the spatiotemporal conditions of human experience,%
\footnote{Position and orientation are in an obvious sense visualizable. Linear and angular momentum derive their meanings from the symmetry properties of space or the invariant behavior of closed systems under translations and rotations, while energy derives its meaning from the uniformity of time or the invariant behavior of closed systems under time translations. Causality and interaction, too, are in obvious ways related to space and time.}
this implies
\begin{quote}
that we have no concepts of the understanding and hence no elements for the cognition of things except insofar as an intuition can be given corresponding to these concepts, consequently that we can have cognition of no object as a thing in itself, but only insofar as it is an object of sensible intuition, i.e. as an appearance; from which follows the limitation of all even possible speculative cognition of reason to mere objects of \emph{experience}. [CPR\,115, original emphasis]
\end{quote}
By placing the subject matter of empirical science squarely into the context of human experience, Kant dispelled many qualms that had been shared by thinkers at the end of the 18th century---qualms about the objective nature of geometry, about the purely mathematical nature of Newton's theory, about the unintelligibility of action at a distance, and about Galileo's principle of relativity. 

Concerning the laws of geometry, which apply to objects constructed by us in the space of our imagination, the question was why they should also apply to the physical world. Kant's answer was that they apply to objects perceived as well as to objects imagined because visual perception and visual imagination share the same space.%
\footnote{\label{note:Kant1}It is noteworthy that Kant's argument applies, not to Euclidean geometry specifically, even though it was the only geometry known in Kant's time, but to geometry in general, and thus to whichever geometry is best suited to formulating the laws of physics. It has even been said that Kant's theory of science set in motion a series of re-conceptualizations of the relationship between geometry and physics that eventuated in Einstein's theories of relativity.\cite{Friedman2009}}
As to the mathematical nature of Newtonian mechanics, it was justified, not by the Neo-Platonic belief that the book of nature was written in mathematical language, but by its being a precondition of scientific knowledge. What makes it possible to conceive of appearances as aspects of an objective world is the mathematical regularities that obtain between them. Newton's refusal to explain action at a distance was similarly justified, inasmuch as the only intelligible causality available to us consists in lawful mathematical relations between phenomena: for the Moon to be causally related to the Earth is for the Moon to stand in a regular mathematical relation to the Earth. As to the principle of relativity, ditto: lawful mathematical relations only exist between phenomena, and thus only between objects or objective events, but never between a particular phenomenon and space or time itself.%
\footnote{\label{note:Kant2}Here, too, it would be an anachronism to argue that Kant singled out Galilean relativity, even though it was the only relativity known in his time. His argument holds for every possible principle of relativity, including Einstein's.}

For this remarkable achievement there was a price to be paid. To preserve the objectivity of science, it must be possible to think of phenomena as appearances of things in themselves: 
\begin{quote}
even if we cannot \emph{cognize} these same objects as [i.e., \emph{know} them to be] things in themselves, we at least must be able to \emph{think} them as things in themselves. For otherwise there would follow the absurd proposition that there is an appearance without anything that appears. [CPR\,115, original emphasis]
\end{quote}
In other words, we must be able to \emph{decontextualize} nature, to free it from the context of human experience, to forget that it is the product of a synthesis achieved by the experiencing subject. The price to be paid was that we must ignore the transcendent reality which affects us in such a way that we have the impressions that we do, and that we are able to organize our impressions into objects that change and interact with each other in accordance with laws of nature.

\section{Kant and the quantum theory}\label{sec.KantQT}
By the time quantum mechanics came along, scientists and philosophers alike had realized that renouncing ontological prejudices and sticking to operationally definable notions was the safest way to arrive at reliable knowledge. At the same time classical physics, still deemed eminently successful, appeared to support a realistic interpretation. What made it possible to reconcile these opposing tendencies was Kant's transcendental philosophy. It offered an ingenious way to go on talking in realist language about, e.g., electromagnetic waves propagating in vacuum, while disavowing ontological inclinations. Kant's transcendental philosophy was therefore widely considered to be tightly linked with classical physics, and to make the latter philosophically acceptable. When classical physics failed to account for such data as atomic spectra, the obvious conclusion was that Kant's philosophy fared no better than na\"{\i}ve realism.\cite{dEspagnat2010}

And indeed, many of Kant's claims appeared to be contradicted by quantum mechanics. There was his principle of thoroughgoing determination, ``according to which, among all possible predicates of things, insofar as they are compared with their opposites, one must apply to it.''[CPR\,553] In direct contradiction to this principle, the properties of atomic systems came to be regarded as possessing values only if (and when) two conditions were satisfied: a set of possible values was defined by an experimental arrangement, and an actual value was indicated. Then there was the necessary and universal truth of \emph{a priori} propositions such as ``the law of the connection of cause and effect''[CPR\,304], established by Kant as preconditions of the possibility of organizing sense-impressions into objects. Yet in the newly discovered quantum domain, there were no sense-impressions waiting to be organized into objects. Neither was it possible to conceive of an atom as a nexus of sense-impressions, nor did atoms satisfy Kant's \emph{a priori} laws. In particular, as was stressed by Schr\"odinger, 
\begin{quote}
Atoms---our modern atoms, the ultimate particles---must no longer be regarded as identifiable individuals. This is a stronger deviation from the original idea of an atom than anybody had ever contemplated. We must be prepared for anything. \cite[p.~162]{SchrNGSH}
\end{quote}
Niels Bohr, seeing Kant as arguing not only for the necessary validity but also the unlimited range of classical concepts, could not but regard his own complementarity interpretation of the quantum formalism as an alternative to Kant's theory of science. And yet---just as Kant did not argue for the universal validity of Euclidean geometry \emph{in particular} (see Note~\ref{note:Kant1}), nor for Galilean relativity \emph{in particular} (see Note~\ref{note:Kant2}), so his arguments did not establish that the range of classical concepts was unlimited. As Kant's arguments had merely established the validity of whichever geometry (and whichever principle of relativity) was the most convenient, so they merely established the necessary validity of classical concepts \emph{as long as} one was dealing with the organization of sense-impressions into objects (which he \emph{assumed} was always the case). Bohr realized that in the new field of experience opened up by the quantum theory one was not \emph{only} dealing with the organization of sense-impressions into objects, and that, consequently, the range of classical concepts was limited---that it did not extend to quantum \emph{systems} but only to quantum \emph{phenomena}. Apart from that,  Bohr established the indispensability of classical concepts in dealing with quantum phenomena by the very same arguments by which Kant had established it for classical phenomena (i.e., for sense-impressions that allow themselves to be organized into objects). Here is Bernard d'Espagnat\cite{dEspagnat2010} on the relation between Kant and contemporary physics:
\begin{quote}
It is true that contemporary physics forces us to give up \dots\ significant, although non central, elements of Kant's thinking. But it more than compensates this blow by practically compelling us to adopt the idea that was, in fact, at the very core of Kantism and constitutes its truly original contribution to philosophical thinking, to wit, the view that things and events, far from being elements of a ``reality per se,'' are just phenomena, that is, elements of our experience.
\end{quote}
Kant did not anticipate the possibility of  an empirical knowledge that, while being obtained \emph{by means of} sense-impressions organized into objects, was not a knowledge \emph{of} sense-impressions organized into objects. Bohr realized that quantum mechanics was that kind of knowledge. He completely agreed with Kant that what is inaccessible to our senses cannot be expected to conform to the spatiotemporal conditions of human experience, and therefore cannot be expected to accord with concepts that owe their meanings to these conditions. 

\section{Niels Bohr}\label{sec.Bohr}
{\leftskip\parindent{It is often said that a work of genius resists categorization. If so, Bohr's philosophical viewpoint easily passes this criterion of greatness. Surely this is one of the reasons for the commonplace complaints over Bohr's ``obscurity.''\par\hfill--- Henry J.\ Folse}\cite{Folse94}\par
\medskip\noindent
{As a philosopher Niels Bohr was either one of the great visionary figures of all time, or merely the only person courageous enough to confront head on, whether or not successfully, the most imponderable mystery we have yet unearthed.\hfill\hbox{--- N. David Mermin}}\cite{MerminBoojums}\par}
\bigskip\noindent
``Without sensibility no object would be given to us,'' Kant wrote [CPR\,193], ``and without understanding none would be thought.'' Bohr could not have agreed more, insisting as he did that meaningful physical concepts have not only mathematical but also visualizable content. Such concepts are associated with pictures, like the picture of a particle following a trajectory or the picture of a wave propagating in space. In the classical theory, a single picture could accommodate all of the properties a system can have. When quantum theory came along, that all-encompassing picture fell apart. Unless certain experimental conditions obtained, it was impossible to picture the electron as following a trajectory (which was nevertheless a routine presupposition in setting up Stern-Gerlach experiments and in interpreting cloud-chamber photographs), and there was no way in which to apply the concept of position. And unless certain other, incompatible, experimental conditions obtained, it was impossible to picture the electron as a traveling wave (which was nevertheless a routine presupposition in interpreting the scattering of electrons by crystals), and there was no way in which to apply the concept of momentum.

If the visualizable content of physical concepts cannot be described in terms of compatible pictures, it has to be described in terms of something that can be so described, and what can be so described are \emph{quantum phenomena}. The definite definition of  a quantum phenomenon is contained in the following passage:
\begin{quote}
[A]ll unambiguous interpretation of the quantum mechanical formalism involves the fixation of the external conditions, defining the initial state of the atomic system concerned and the character of the possible predictions as regards subsequent observable properties of that system. Any measurement in quantum theory can in fact only refer either to a fixation of the initial state or to the test of such predictions, and it is first the combination of measurements of both kinds which constitutes a well-defined phenomenon. [BCW7:\,312]%
\footnote{These references are to the volumes of the \emph{Collected Works} of Niels Bohr.\cite{BCW}}
\end{quote}
Today, Bohr is mostly known for his insistence on the necessity of using classical concepts, for attributing this necessity to the need to communicate to others ``what we have done and what we have learned'' [BCW7: 273, 331, 349, 390, 418],  and for the thesis that ``the specification of [the whole experimental arrangement] is imperative for any well-defined application of the quantum-mechanical formalism''.\cite{BohrSchilpp} The conceptual links between these demands, however, belong to a fabric of thought that is not widely known. In the remainder of this section and the three sections that follow, an attempt is made at an outline of the overarching framework of Bohr's thought.

In a 1922 letter to his philosophical mentor Harald H{\o}ffding, Bohr wrote:
\begin{quote}
my personal opinion is that these difficulties are of such a kind that they hardly allow us to hope, within the world of atoms, to implement a description in space and time of the kind corresponding to our usual sensory images. [BCW10:\,513]
\end{quote}
In each of the following quotes, all from 1929, Bohr refers to space and time as our ``forms of perception'':
\begin{quote}
[T]he very recognition of the limited divisibility of physical processes \dots\ has justified the old doubt as to the range of our ordinary forms of perception when applied to atomic phenomena. [BCW6:\,209]

\medskip[W]e can hardly escape the conviction that in the facts which are revealed to us by the quantum theory and lie outside the domain of our ordinary forms of perception we have acquired a means of elucidating general philosophical problems. [BCW6:\,217]

\medskip This limitation [of our forms of perception] is brought to light by a closer analysis of the applicability of the basic physical concepts in describing atomic phenomena. [BCW6:\,242]

\medskip[W]e must remember, above all, that, as a matter of course, all new experience makes its appearance within the frame of our customary points of view and forms of perception. [BCW6:\,279]

\medskip[W]e must not forget that, in spite of their limitation, we can by no means dispense with those forms of perception which colour our whole language and in terms of which all experience must ultimately be expressed.\hfill\break [BCW6:\,283]

\medskip[T]he difficulties concerning our forms of perception, which arise in the atomic theory\dots, may be considered as an instructive reminder of the general conditions underlying the creation of man’s concepts. [BCW6:\,293]

\medskip[A]ll our ordinary verbal expressions bear the stamp of our customary forms of perception, from the point of view of which the existence of the quantum of action is an irrationality. Indeed, in consequence of this state of affairs, even words like ``to be'' and ``to know'' lose their unambiguous meaning. [BCW6:\,297]
\end{quote}
Today the task of making sense of the quantum theory is widely seen as one of grafting a metaphysical narrative onto a mathematical formalism, in a language that is sufficiently vague philosophically to be understood by all and sundry. For Bohr, as also for Heisenberg and Pauli, the real issues lay deeper. They judged that the conceptual difficulties posed by the quantum theory called in question the general framework of thought that had evolved in Germany beginning with Kant. If (i)~space and time are but our forms of perception, if (ii)~physical concepts derive their meanings from different aspects of these forms (e.g., localizability and homogeneity or invariance under translations), and if (iii)~the facts revealed to us by the quantum theory lie outside the domain of our ordinary forms of perception (in other words, if they are inaccessible to sensory perception), then these facts cannot be expected to be expressible by the physical concepts at our disposal. How, then, can they be expressed, and how can this be done without compromising the objectivity of the theory? Bohr's answer in a nutshell:
\begin{quote}
the decisive point is that the physical content of quantum mechanics is exhausted by its power to formulate statistical laws governing observations obtained under conditions specified in plain language. [BCW10:\,159] 
\end{quote}
By developing the mathematical part of the quantum theory into an autonomous formal language,  von Neumann\cite{vN1932} transformed the theory into a mathematical formalism in search of a physical interpretation. Transmogrifying a probability algorithm---the state vector---into an evolving physical state, adopting the eigenvalue-eigenstate link, and modeling measurements as two-stage processes (``pre-measure\-ment'' followed by ``objectification''), he gave rise to what has been appropriately called ``the disaster of objectification'' by van Fraassen.\cite{vF1990} This is how quantum mechanics became ``the great scandal of physics'',\cite{Wallace2008} ``the silliest'' of all the theories proposed in the 20th century,\cite{Kaku95} and a theory that ``makes absolutely no sense''.\cite{Penrose86} A distinction is made between  the ``bare quantum formalism,'' which is regarded as ``an elegant piece of mathematics \dots\ prior to any notion of probability, measurement etc.,'' and the ``quantum algorithm,'' which is looked upon as ``an ill-defined and unattractive mess''\cite{Wallace2008}.%
\footnote{In reality there is no such thing as a \emph{bare} quantum formalism. Every single axiom of any axiomatization of the theory only makes sense as a feature of a probability calculus.\cite{Mohrhoff-QMexplained,Mohrhoff-book20.2} The distinction between a bare quantum formalism and a quantum algorithm is as illegitimate as the distinction between the ``easy'' problems of consciousness and the ``hard'' one.\cite{Chalmers95,Lowe95} Both distinctions are rooted in the obsolescent mode of thinking that is known as ``physicalism.''}
``Measurement'' has become the unmentionable M-word of physics.\cite{Bell90} And Bohr, of all people, often gets blamed for this sorry state of affairs!%
\footnote{Even by QBists: ``The Founders of quantum mechanics were already aware that there was a problem. Bohr and Heisenberg dealt with it by emphasizing the inseparability of the phenomena from the instruments we devised to investigate them\dots. Being objective and independent of the agent using them, instruments miss the central point of QBism, \emph{giving rise to the notorious measurement problem}, which has vexed physicists to this day''.\cite[emphasis added]{FMS2014} In actual fact, it was von Neumann who gave rise to this problem. For Bohr there was ``no new observational problem''[BCW10:\,212] because we are doing what we have always done: setting up experiments and reporting their results.}

If measurements and plain language played pivotal roles in Bohr's writings, it was to ensure the objectivity of the new theory. When Bohr realized that his references to ``sensory images'' and ``forms of perception'' rather contributed to undermining his efforts in that direction, Bohr replaced these expressions by ``quantum phenomena'' and ``experimental arrangements.'' I owe this observation to the editor of the two volumes of Bohr's \emph{Collected Works} that deal specifically with the foundations of quantum physics:
\begin{quote}
when the phrase ``forms of perception'' was replaced by ``experimental arrangement'', ``the objectivity of physical observations'' could be stressed without the somewhat bewildering addition that it could be ``particularly suited to emphasize the subjective character of all experience''.\cite{Kalckar96} 
\end{quote}
While the business of physics was ``the development of methods for ordering and surveying human experience,'' this was to be done ``in a manner independent of individual subjective judgement and therefore objective in that sense, that it can be unambiguously communicated in the common human language'' [BCW10: 157--158]:
\begin{quote}
To clarify this point [whether we are concerned with a complete description of natural phenomena], it was indeed necessary to examine what kind of answers we can receive by so to say putting questions to nature in the form of experiments. In order that such answers may contribute to objective knowledge, independent of subjective judgement, it is an obvious demand that the experimental arrangement as well as the recording of observations be expressed in the common language, developed for our orientation in the surroundings. [BCW10:\,212]
\end{quote}
At one time Heisenberg\cite{Heisenberg1935} drew a dividing line between ``the apparatus which we\dots, in a way, treat as part of ourselves,'' and ``the physical systems we wish to investigate.'' Pauli likewise thought that it was ``allowed to consider the instruments of observation as a kind of prolongation of the sense organs of the observer'' [BCW10:\,564]. Bohr would have none of this. The observed had to be detached from the observer (rather than the other way round), and there was only one way to do this: to take the means of observation, rather than the system observed, for what was \emph{actually} observed, what was \emph{directly} accessible to our senses, and what therefore was amenable to communication using words and concepts we can understand. The dividing line was to be drawn, not between the apparatus as part of ourselves and the object of investigation, but between our observing selves and the observed apparatus. What could not be separated from the object of investigation was not the subject, which remained the same detached observer it had been before quantum physics came along, but the means of investigation. And this was not a matter of choice, for without the apparatus not only did the object of investigation lack properties but, in fact, there was no object of investigation.

\section{Quantum systems or quantum phenomena?}\label{sec.QsysQpha}
Where there are no sense-impressions waiting to be organized into objects, there are no objects. Bohr's emphatic rejection of the familiar language of objects when dealing with ``the facts which are revealed to us by the quantum theory'' cannot be overemphasized:
\begin{quote}
\label{p:ambiguity}The unaccustomed features of the situation with which we are confronted in quantum theory necessitate the greatest caution as regards all questions of terminology. Speaking, as is often done, of disturbing a phenomenon by observation, or even of creating physical attributes to objects by measuring processes, is, in fact, liable to be confusing, since all such sentences imply a departure from basic conventions of language which, even though it sometimes may be practical for the sake of brevity, can never be unambiguous. It is certainly far more in accordance with the structure and  interpretation of the quantum mechanical symbolism, as well as with  elementary epistemological principles, to reserve the word ``phenomenon'' for the comprehension of the effects observed under given  experimental conditions. [BCW7:\,316]
\end{quote}
If there is no object to be disturbed by a measurement, if even the dichotomy of  objects and attributes created for them by measuring processes is unwarranted, then it is not just the measured property but the quantum system itself that is constituted by the experimental conditions under which it is observed. 

More recently this point was forcefully made by Brigitte Falkenburg in her commendable monograph \emph{Particle Metaphysics}:
\begin{quote}
[O]nly the experimental context (and our ways of conceiving of it in classical terms) makes it possible to talk in a sloppy way of \emph{quantum objects}\dots. Bare quantum ``objects'' are just bundles of properties which underlie superselection rules and which exhibit non-local, acausal correlations\dots. They seem to be Lockean empirical substances, that is, collections of empirical properties which constantly go together. However, they are only individuated by the experimental apparatus in which they are measured or the concrete quantum phenomenon to which they belong\dots. They can only be individuated as context-dependent quantum \emph{phenomena}. Without a given experimental context, the reference of quantum concepts goes astray. In this point, Bohr is absolutely right up to the present day. \cite[pp. 205--206, original emphases]{Falkenburg2007}
\end{quote}
A similar conclusion was reached by Ole Ulfbeck and Aage Bohr,\cite{UlfbeckBohr} for whom ``there is no longer a particle passing through the apparatus and producing the click. Instead, the connection between source and counter is inherently non-local.'' While ``clicks can be classified as electron clicks, neutron clicks, etc., \dots there are no electrons and neutrons on the spacetime scene.'' Hence ``there is no wave function for an electron or a neutron but a wave function for electron clicks and neutron clicks, etc.'' What makes it seem as if there are electrons and neutrons is the existence of conservation laws, which govern patterns of clicks. If a ``+\,click'' is always followed by a ``+\,click'' we seem to have the right to infer the continued existence of a ``+\,particle,'' but if a ``+\,click'' can also be followed by two ``+\,clicks'' and a ``$-$\,click'' or by three ``+\,clicks'' and two ``$-$\,clicks'' then, as Schr\"odinger put it, ``we must be prepared for anything.''

\section{Objectivity, ordinary language, and classical concepts}\label{sec.OLCC}
Presently (July 2019) a combined Google search for ``Bohr'' and ``classical language'' (the latter term including the quotes) yields more than 5,000 results. A search for ``Bohr'' and ``language of classical physics'' yields nearly 25,000 results. By contrast, searching the 13 volumes of the \emph{Complete Works} of Niels Bohr does not yield a \emph{single} occurrence of either ``classical language'' or ``language of classical physics.'' While Bohr insisted on the use of classical \emph{concepts} for describing quantum phenomena,%
\footnote{Sometimes Bohr refers instead to ``elementary physical concepts'': ``all subjectivity is avoided by proper attention to the circumstances required for the well-defined use of elementary physical concepts'' [BCW7:\,394].}
the \emph{language} on the use of which he insisted was ``ordinary language'' [BCW7:\,355], ``plain language'' [BCW10:\,159], the ``common human language'' [BCW10:\,157--158], or the ``language common to all'' [10:\,xxxvii].%
\footnote{Jan Faye\cite{Faye_Bohr_Darwin} has argued that ``Bohr was not a transcendentalist in his insistence on the use of classical concepts. Instead he had a naturalistic attitude to how common language came about.'' Certain passages from the Bohr canon can be adduced in support of this claim, e.g., when Bohr insists on the use of the ``common language developed for our orientation in the surroundings'' [BCW10:\,212], or when he points out that in classical physics the goal of an objective description is secured by the circumstance that such descriptions are ``based on pictures and ideas embodied in common language, adapted to our orientation in daily-life events'' [BCW10:\,276]. I do not think, however, that transcendentalist and naturalistic attitudes are mutually exclusive, nor that Bohr's motivation for insisting on the use of classical concepts was primarily naturalistic.}

To represent the content of my experience as objective, I do not need to represent it as a system of objects located in space and changing with time in such a way that they can be re-identified and compared, as Kant had taught, but I need to be able to refer to such objects, and for this I need ordinary language and classical concepts. Ordinary human language uses words we can all understand, inasmuch as their meanings are rooted in what is common to us, i.e., the spatiotemporal structure of human experience and the logic of human thought or the structure of human language. This includes the classical concepts. Bohr often referred to ordinary language and classical concepts (of equivalents thereof) in the same breath. What is required is not classical physics but only the \emph{terminology} of classical physics: ``all well-defined experimental evidence, \emph{even if it cannot be analysed in terms of classical physics}, must be expressed in ordinary language'' [BCW7:\,355; emphasis added], i.e., ``plain language suitably refined by the usual physical terminology'' [BCW7:\,390] or ``conveniently supplemented with terminology of classical physics'' [BCW10:\,277].

While classical concepts and ordinary language are necessarily used in both classical physics and quantum physics, in quantum physics their use is restricted to the domain of re-identifiable objects with intrinsic attributes, which in classical physics is all there is. Quantum physics reveals a domain to which neither ordinary language nor classical concepts can legitimately be applied---an intrinsically unspeakable domain which becomes speakable only indirectly, via an experimental context. So: objectivity $\Rightarrow$ ordinary language and classical concepts $\Rightarrow$ contextuality: 
\begin{quote}
By objectivity we understand a description by means of a language common to all (quite apart from the differences in languages between nations) in which people may communicate with each other in the relevant field. [BCW10:\,xxxvii]

\medskip From a logical standpoint, we can by an objective description only understand a communication of experience to others by means of a language which does not admit ambiguity as regards the perception of such communications. In classical physics, this goal was secured by the circumstance that, apart from unessential conventions of terminology, the description is based on pictures and ideas embodied in common language, adapted to our orientation in daily-life events. [BCW10:\,276]

\medskip Faced with the question of how under such circumstances we can achieve an objective description, it is decisive to realize that however far the phenomena transcend the range of ordinary experience, the description of the experimental arrangement and the recording of observations must be based on common language. [BCW10:\,158]
\end{quote}
One day during tea at his institute, Bohr was sitting next to Edward Teller and Carl Friedrich von Weizs\"acker. Von Weizs\"acker\cite{vW_Structure} recalls that when Teller suggested that ``after a longer period of getting accustomed to quantum theory we might be able after all to replace the classical concepts by quantum theoretical ones,'' Bohr listened, apparently absent-mindedly,  and said at last: ``Oh, I understand. We also might as well say that we are not sitting here and drinking tea but that all this is merely a dream.'' If we are dreaming, we are unable to tell others what we have done and what we have learned. Therefore
\begin{quote}
it would be a misconception to believe that the difficulties of the atomic theory may be evaded by eventually replacing the concepts of classical physics by new conceptual forms. \dots the recognition of the limitation of our forms of perception by no means implies that we can dispense with our customary ideas or their direct verbal expressions when reducing our sense impressions to order. [BCW6:\,294]
\end{quote}
Or, as Heisenberg put it,\cite[p.~56]{Heisenberg_PP} ``[t]here is no use in discussing what could be done if we were other beings than we are.''%
\footnote{Heisenberg thought it possible that the forms of perception of other beings, and hence their concepts, could be different from ours: ours ``may belong to the species `man,' but not to the world as independent of men''.\cite[p.~91]{Heisenberg_PP}} 
Bohr's claim that the ``classical language'' (i.e., plain language supplemented with terminology of classical physics) was indispensable, has also been vindicated by subsequent developments in particle physics:
\begin{quote}
This [claim] has remained valid up to the present day. At the \emph{individual} level of clicks in particle detectors and particle tracks on photographs, all measurement results have to be expressed in classical terms. Indeed, the use of the familiar physical quantities of length, time, mass, and momentum-energy at a subatomic scale is due to an extrapolation of the language of classical physics to the non-classical domain.\cite[p.~162]{Falkenburg2007}
\end{quote}
 
\section{Irreversible amplification?}\label{sec.amplification}
If the terminology of quantum phenomena is used consistently, then nothing---at any rate, nothing we know how to think about---happens between ``the fixation of the external conditions, defining the initial state of the atomic system concerned'' and ``the subsequent observable properties of that system'' [BCW7:\,312]. Any story purporting to detail a course of events in the interval between a system preparation and a subsequent observation is inconsistent with ``the essential wholeness of a quantum phenomenon,'' which ``finds its logical expression in the circumstance that any attempt at its subdivision would demand a change in the experimental arrangement incompatible with its appearance'' [BCW10:\,278]. What, then, are we to make of the following passages [emphases added]?
\begin{quote}
[E]very well-defined atomic phenomenon is closed in itself, since its observation implies a permanent mark on a photographic plate \emph{left by the impact of an electron} or similar recordings \emph{obtained by suitable amplification devices of essentially irreversible functioning}. [BCW10:\,89]

\medskip Information concerning atomic objects consists solely in the marks they make on these measuring instruments, as, for instance, a spot \emph{produced by the impact of an electron on a photographic plate} placed in the experimental arrangement. The circumstance that such marks are \emph{due to irreversible amplification effects} endows the phenomena with a peculiarly closed character pointing directly to the irreversibility in principle of the very notion of observation. [BCW10:\,120]

\medskip In this connection, it is also essential to remember that all unambiguous information concerning atomic objects is derived from the permanent marks---such as a spot on a photographic plate, \emph{caused by the impact of an electron}---left on the bodies which define the experimental conditions. Far from involving any special intricacy, the\emph{ irreversible amplification effects on which the recording of the presence of atomic objects rests} rather remind us of the essential irreversibility inherent in the very concept of observation. [BCW7:\,390; BCW10:\,128]
\end{quote}
If a well-defined atomic phenomenon is closed, how can there be something between the fixation of the external conditions and a permanent mark on a photographic plate? Does not the interposition of the impact of an electron and of a subsequent amplification effect amount to a subdivision of the phenomenon in question? 

Ulfbeck and (Aage) Bohr\cite{UlfbeckBohr} have shed light on this issue. Like Kant and subsequently (Niels) Bohr, they view space and time as ``a scene established for the ordering of experiences.'' Clicks in counters are ``events in spacetime, belonging to the world of experience.'' All physical phenomena are described in terms of variables that have values at all times, and each variable of this kind belongs to an object that is present on the spacetime scene---a ``classical'' or ``macroscopic'' object in customary parlance. The matrix variables of quantum mechanics are ``of an entirely novel type.'' A click is an event ``by which a matrix variable manifests itself on the spacetime scene, without entering this scene.'' The key to resolving the issue at hand is that this event---the click---has an ``onset'':
\begin{quote}
[A] click is distinguished by the remarkable property of having an ``onset,'' a beginning from which the click evolves as a signal in the counter. The onset, thus, has no precursor in spacetime and, hence, does not belong to a chain of causal events. In other words, the onset of the click is not the effect of something, and it has no meaning to ask how the onset occurred\dots. The notion that a particle entered the counter, therefore, becomes inappropriate, and it is likewise inappropriate to state that the particle produced the click\dots. [T]he downward path from macroscopic events in spacetime, which in standard quantum mechanics continues into the regime of the particles, does not extend beyond the onsets of the clicks\dots. [T]he occurrence of genuinely fortuitous clicks, coming by themselves, is recognized as the basic material that quantum mechanics deals with\dots. The theory of what takes place in spacetime is, therefore, inherently non-local\dots. Thus, the wave function enters the theory not as an independent element, but in the role of encoding the probability distributions for the clicks, which is the content of the theory.
\end{quote}
In the conventional/orthodox picture, a particle impinges on the counter and produces a chain of processes leading to the click. For Ulfbeck and Bohr, ``there is no incident particle, and the steps in the development of the click, envisaged in the usual picture, are not events that have taken place on the spacetime scene.''

To (Niels) Bohr, ``the quantum-mechanical formalism \dots\ represents a purely symbolic scheme permitting only predictions \dots\ as to results obtainable under conditions specified by means of classical concepts'' [BCW7: 350--351]. Because ``the physical content of quantum mechanics is exhausted by its power to formulate statistical laws governing observations obtained under conditions specified in plain language'' [BCW10:\,159], a quantum phenomenon has a mathematical or statistical part and a part that relates the same to experience. The irreversible amplification effects belong to the former. The unmediated step from the so-called source to the onset of the click, and the subsequent unmediated steps in the development of the click, are steps in a gazillion of alternative sequences of possible outcomes of {unperformed} measurements, and unperformed measurements have no effect on the essential wholeness of a quantum phenomenon.

In the following passages [emphases added], Bohr appears to argue (rather unnecessarily) that the quantum features involved in the atomic constitution of the measurement apparatus (or the statistical element in its description) can be neglected \emph{because} the relevant parts of the apparatus are sufficiently large and heavy.
\begin{quote}
In actual experimentation this demand [that the experimental arrangement as well as the recording of observations be expressed in the common language] is met by the specification of the experimental conditions by means of bodies like diaphragms and photographic plates \emph{so large and heavy that the statistical element in their description can be neglected}. The observations consist in the recording of permanent marks on these instruments, and the fact that the amplification devices used in the production of such marks involves essentially irreversible processes presents no new observational problem, but merely stresses the element of irreversibility inherent in the definition of the very concept of observation. [BCW10:\,212]

\medskip In actual physical experimentation this requirement [that we must employ common language to communicate what we have done and what we have learned by putting questions to nature in the form of experiments] is fulfilled by using as measuring instruments rigid bodies like diaphragms, lenses, and photographic plates sufficiently large and heavy to allow an account of their shape and relative positions and displacements without regard to any quantum features inherently involved in their atomic constitution\dots. The circumstance that [recordings of observations like the spot produced on a photographic plate by the impact of an electron] involve essentially irreversible processes presents no special difficulty for the interpretation of the experiments, but rather stresses the irreversibility which is implied in principle in the very concept of observation. [BCW10:\,165]
\end{quote}
How can the size or weight of a measuring device justify
\begin{itemize}
\item[---] the irreversibility in principle of the very notion of observation\hfill\break [BCW10:\,120],
\item[---] the essential irreversibility inherent in the very concept of observation\hfill\break[BCW7:\,390; BCW10:\,128],
\item[---] the irreversibility which is implied in principle in the very concept of observation [BCW10:\,165], or
\item[---] the element of irreversibility inherent in the definition of the very concept of observation [BCW10:\,212]?
\end{itemize}
The only irreversibility that can justify the irreversibility of observations is the incontestable irreversibility of human sensory experience. For Bohr, ``the emphasis on the subjective character of the idea of observation [was] essential'' [BCW10:\,496]. If the description of atomic phenomena nevertheless had ``a perfectly objective character,'' it was ``in the sense that no explicit reference is made to any \emph{individual} observer and that therefore \dots\ no ambiguity is involved in the communication of information'' [BCW7:\,390, emphasis added]. It was never in the sense that no explicit reference was made to the community of communicating observers or to the incontestable irreversibility of their experiences. It is only if one wants to disavow \emph{any} reference to experience that one may have to invoke something like a sufficiently large or heavy apparatus. Bohr may have intended in this way to appease the na\"{\i}ve realistic inclinations of lesser minds, but it certainly does not characterize his  own philosophical stance.

\section{QBism, language, and the external world}\label{sec.QBLEW}
There can be no doubt that significant progress has been made during the roughly four decades between the passing of Niels Bohr and the advent of QBism. We now have a congeries of complex, sophisticated, and astonishingly accurate probability algorithms---the standard model%
\footnote{ ``Standard model is a grotesquely modest name for one of humankind's greatest achievements''.\cite{Wilczek2008}}%
---and we are witnessing rapid growth in the exciting fields of quantum information and quantum technology. By contrast, the contemporaneous progress in quantum theory's philosophical foundations mainly consists in finding out what does not work. This includes the various attempts that were made to transmogrify statistical correlations between observations into physical processes taking place between or giving rise to observations.

Whereas a quantum state exists in a Hilbert space, \emph{we} live in a 3D space, at least in the sense that it frames our experience of an external world. Hilbert space ``knows'' nothing about a world of objects localized in a 3D~space. \emph{We} do. To make the abstract Hilbert space relevant to our experience, we represent it as a space of wave functions defined on a configuration space and evolving in time. This configuration space and this time are not part of the aforesaid ``elegant piece of mathematics'' (the so-called bare quantum formalism); they belong to the ``ill-defined and unattractive mess'' (the so-called quantum algorithm) to which that ``elegant piece of mathematics'' owes its physical meaning and relevance.

As a system of \emph{formal} propositions, quantum mechanics allows us to call a self-adjoint operator ``elephant'' and a spectral decomposition ``trunk.'' This makes it possible to prove a theorem according to which every elephant has a trunk. Either the words ``time'' and ``space'' mean as little as the word ``elephant'' does when it is used in this way, or they owe their meanings to human experience, in which case wave functions encapsulate correlations between experiences. Self-adjoint operators on a Hilbert space do not become observables by calling them ``observables,'' any more than they become elephants by calling them ``elephants.'' They come to be associated with observables when they are seen as tools for assigning probabilities to the possible outcomes of measurements, which do not happen in a Hilbert space. The real business of interpreting the quantum formalism is to situate it in the experiential framework in which ``time'' and ``space'' make sense, not to use these words in ways that divest them of their meanings.

The great merit of QBism is to put the spotlight back on the role that human experience plays in creating physical theories. If measurements are irreversible and outcomes definite, it is because our experiences are irreversible and definite. This at once disposes of the disaster of objectification. Bohr could have said the same, and arguably did, but in such elliptic ways that the core of his message has been lost or distorted beyond recognition. The fundamental difference between Bohr and QBism is that one was writing \emph{before} interpreting quantum mechanics became a growth industry, while the other emerged in reaction to the ever-growing number of futile attempts at averting the disaster of objectification.

To make the centrality of human experience truly stick, QBism emphasizes the \emph{individual} subject. It is not \emph{we} who experience the world. At first the experience is not \emph{ours}; it is \emph{yours} and \emph{mine}. It \emph{becomes} ours, and the process by which it becomes ours is \emph{communication}:
\begin{quote}
What is real for an agent rests entirely on what that agent experiences, and different agents have different experiences. An agent-dependent reality is constrained by the fact that different agents can communicate their experience to each other, limited only by the extent that personal experience can be expressed in ordinary language. Bob's verbal representation of his own experience can enter Alice's, and vice-versa. In this way a common body of reality can be constructed.\cite{FMS2014}
\end{quote}
This was also Schr\"odinger's take.\cite{SchrWhatIsReal} Here is how he set up the question:  
\begin{quote}
I get to know the external world through my sense-perceptions. It is only through them that such knowledge flows into me; they are the very material out of which I construct it. The same applies to everyone else. The worlds thus produced are, if we allow for differences in perspective, etc., very much the same, so that in general we use the singular: world. But because each person's sense-world is strictly private and not directly accessible to anyone else, this agreement is strange; what is especially strange is how it is established\dots. This is a valid question: how do we come to know of this general agreement between two private worlds, when they admittedly are private and always remain so? Direct comparison does not help, for there is none. It is absolutely necessary that we should start by being deeply troubled by the monstrous character of this state of affairs, if we are to treat with indulgence the inadequate attempts that have been made to explain it. 
\end{quote}
So what establishes the correspondence ``between the content of any one sphere of consciousness and any other, so far as the external world is concerned''?
\begin{quote}
What does establish it is \emph{language}, including everything in the way of expression, gesture, taking hold of another person, pointing with one's finger and so forth, though none of this breaks through that inexorable, absolute division between spheres of consciousness.
\end{quote}
As far as the external world is concerned, it thus seems that Bohr, Schr\"odinger, and the QBists are all on the same page. They all agree that experience provides the material from which we construct ``a common body of reality,'' and that language is the means by which we construct it. 

But here I must respond to something I strongly disagree with. Mermin\cite{MerminQBnotCop} claims that ``[o]rdinary language comes into the QBist story in a more crucial way than it comes into the story told by Bohr.'' He supports this by  saying that ``measurement outcomes in QBism are necessarily classical, in a way that has nothing to do with language,'' implying that for Bohr the classicality of measurement outcomes had something to do with language. I maintain that it had not. For QBism as for Bohr, measurement outcomes are definite and irreversible because experiences are definite and irreversible. 

Bohr would also agree with Mermin that ``language is the only means by which different users of quantum mechanics can attempt to compare their own private experiences'' (though he might have pointed out that this is true for everyone, not just for users of quantum mechanics). What he might not have agreed with is the idea that each of us first constructs a private external world, and that language comes in only after this is done, as a means of figuring out ``what is common to all our privately constructed external worlds.'' He might have pointed out that one cannot construct a private external world before being in possession of a language providing the concepts that are needed for its construction. 

Arguably nobody has shed more light than Kant on how each of us constructs her or his private external world, assuming that we are in possession of the relevant concepts. In Kant's view, I construct a system of interacting, re-identifiable objects by combining relations that owe their meanings to our ``forms of perception'' with relations that owe their meanings to the logical structure of our thought or the grammatical structure of ``a language common to all.'' Here I made use of phrases from the Bohr canon in order to highlight Bohr's affinity with Kant, and I wrote \emph{our} ``forms of perception'' and \emph{our} thought because we could never communicate with each other and arrive at a common external world if my forms of perception and the logical structure of my thought were different from yours. The essential similarity of the general form of my perceptions and my thoughts with yours is a presupposition that is implicit in every statement about the external world.

It might seem at first that Kant had little to say about how we make the move from privately constructed external worlds to a shared external world. But if my forms of perception are the same as yours, and if the logical structure of my thought is identical to yours, then we use the same concepts in constructing our private external worlds. And then it makes little difference whether we talk to ourselves silently (i.e., construct our private external worlds) or talk to each other loudly (i.e., construct our common external world). Mermin overlooks that language plays the same constitutive role in the construction of his own private external world as it does in the  ``collaborative human effort to find \dots\ a model for what is common to all of our privately constructed external worlds,'' which is the QBist view of science.

\section{Locating the boundary of our common external world}\label{sec.boundary}
John Bell is famous for, among other things, his disapproval of the word ``measurement'' in quantum mechanics textbooks and the ``shifty split of the world into `system' and `apparatus'\,'' thereby entailed:
\begin{quote}
There can be no question---without changing the axioms---of getting rid of the shifty split. Sometimes some authors of `quantum measurement' theories seem to be trying to do just that. It is like a snake trying to swallow itself by the tail. It can be done---up to a point. But it becomes embarrassing for the spectators even before it becomes uncomfortable for the snake.\cite{Bell90} 
\end{quote}
While for Heisenberg the location of the split (a.k.a. the Heisenberg cut) was to some extent arbitrary, for Bohr it was unambiguously determined by the measurement setup.%
\footnote{If the diaphragm is fixed, it is part of the experimental arrangement; if it is moveable, it is part of the system under investigation.\cite{BohrSchilpp} See Camilleri and Schlosshauer\cite{CamilleriSchlosshauer2015} for a discussion of Bohr's and Heisenberg's divergent views on this matter.}
In an attempt to get rid of the shiftiness of the split, QBists put the measurement \emph{outcome} into the mind of an agent and replace the measurement \emph{setup} by any old action taken by the agent: a measurement is an action taken to elicit one of a set of possible experiences, and the outcome of a measurement is the experience elicited by such an action.%
\footnote{While Fuchs and Schack prefer the term ``agent,'' Mermin prefers the term ``user,'' to emphasize that QBists regard quantum mechanics as a ``user's manual''.\cite{MerminQBnotCop}}
Accordingly there are as many splits between the agent-experiencer and the system acted on as there are agent-experiencers, and there is nothing shifty about the splits:
\begin{quote}
Each split is between an object (the world) and a subject (an agent's irreducible awareness of her or his own experience). Setting aside dreams or hallucinations, I, as agent, have no trouble making such a distinction, and I assume that you don't either. Vagueness and ambiguity only arise if one fails to acknowledge that the splits reside not in the objective world, but at the boundaries between that world and the experiences of the various agents who use quantum mechanics.\cite{Mermin_shifty}
\end{quote}
Let us disregard the philosophically ambiguous concept of awareness of one's own experience. The trouble with this approach is that it poses a dilemma. If QBism treats ``all physical systems in the same way, including atoms, beam splitters, Stern-Gerlach magnets, preparation devices, measurement apparatuses, all the way to living beings and other agents'',\cite{FS2015} and if the action taken ``can be anything from running across the street at L'\'{E}toile in Paris (and gambling upon one's life) to a sophisticated quantum information experiment (and gambling on the violation of a Bell inequality)'',\cite{Fuchs_Notwithstanding} then Bohr's crucial insight that the properties of quantum systems are \emph{contextual}---that they are \emph{defined} by experimental arrangements---is lost.  

To preserve this contextuality, Fuchs re-introduces the apparatus (which however does not seem to feature in the act of running across the street at L'\'{E}toile) as a part of the agent:
\begin{quote}
QBism holds with Pauli (and against Bohr) that a measurement apparatus must be understood as an extension of the agent himself, not something foreign and separate. A quantum measurement device is like a prosthetic hand, and the outcome of a measurement is an unpredictable, undetermined ``experience'' shared between the agent and the external system.\cite{Fuchs_Notwithstanding}
\end{quote}
Whereas orthodoxy has it that the experimenter acts on laboratory equipment (i.e., manipulates it with her actual hands), according to Fuchs she acts on the ``external system'' using her prosthetic hand. This brings us to the other horn of the dilemma, for now it is not clear where the apparatus---the prosthetic hand---ends and the rest of the laboratory begins. It appears that one shifty split has been traded for another. 

In Bohr's view, a measurement apparatus serves not only to indicate the possession of a property (by a system) or a value (by an observable) but also, and in the first place, to make a set of properties or values available for attribution to a system or an observable. The sensitive regions of an array of detectors \emph{define} the regions of space in which the system can be found. In the absence of an array of detectors, the regions of space in which the system can be found do not exist. The orientation of a Stern-Gerlach apparatus \emph{defines} the axis with respect to which a spin component is measured.  In the absence of a Stern-Gerlach apparatus, the axis with respect to which a spin component can be up or down does not exist. What physical quantity is defined by running across the street at L'\'{E}toile in Paris?

We cannot describe an object without describing it. For describing an object we need concepts, and if we want to describe a quantum system, we need to make the concepts that are available to us applicable to it. And for this we need experimental arrangements. There is a clear-cut demarcation between the object of investigation and the means of investigation: the means we can describe directly, the object only indirectly, in terms of correlations between what we may do and what we may learn as a result. No such clear-cut demarcation exists between what forms part of the experimenter's prosthetic hand and what does not.

Fuchs' response to this challenge is that the physical extent of the agent is up to the agent:
\begin{quote}
The question is not where does the quantum world play out and the classical world kick in? But where does the agent expect his own autonomy to play out and the external world, with its autonomy and its capacity to surprise, kick in? The physical extent of the agent is a judgment he makes of himself.\cite{Fuchs2Wootters}
\end{quote}
While Mermin places the quantum/classical divide between each individual user’s subjective experience and ``the common external world we have all negotiated with each other'',\cite{MerminQBnotCop} Fuchs places it between the agent-cum-instrument and the rest of the physical world. In other words, the dividing line---wherever the agent chooses to place it---is drawn \emph{in} the objective world, which is precisely what Mermin objects to when he writes that ``[v]agueness and ambiguity only arise if one fails to acknowledge that the splits reside not in the objective world, but at the boundaries between that world and the experiences of the various agents who use quantum mechanics''.\cite{Mermin_shifty}

The single most important message of QBism is that the definiteness of measurement outcomes and the resultant irreversibility of measurements are rooted in the incontestable definiteness and irreversibility of each human individual's sensory experience. QBists seem to believe that they are therefore required to treat all physical systems in the same way, from subatomic particles to measurement apparatuses and on to all agents except the experiencing one. They thus feel compelled to consider the outlandish possibility of ``some amazing quantum interference experiment'' that puts Wigner's friend in a coherent superposition of having experienced two different measurement outcomes. 

In Wigner's scenario,\cite{Wigner61} his's friend ($F$) performs a measurement on a system~$S$ using an apparatus~$A$. Treating $F$ as a quantum system, and treating quantum states as ontic states evolving unitarily between measurement-induced state reductions, Wigner concludes that a reduction of the combined system $S{+}A$ occurs for $F$ when she becomes aware of the outcome, while a reduction of  the combined system $S{+}A{+}F$ occurs for him when he is informed of the outcome by~$F$. This led Wigner to conclude that the theory of measurement was logically consistent only ``so long as I maintain my privileged position as ultimate observer.'' QBism, by contrast, maintains that Wigner's state assignments, which are based on his actual past and possible future experiences, are as valid as his friend's, based as they are on different sets of actual past and possible future experiences. This important point, however, can be made without envisioning Wigner's friend in a coherent superposition of two distinct cognitive states:
\begin{quote}
Wigner's quantum-state assignment and unitary evolution for the compound system are only about his \emph{own} expectations for his \emph{own} experiences should he take one or another action upon the system or any part of it. One such action might be his sounding the verbal question, ``Hey friend, what did you see?,'' which will lead to one of two possible experiences for him. Another such action could be to put the whole conceptual box into some amazing quantum interference experiment, which would lead to one of two completely different experiences for him.\cite{Fuchs_Notwithstanding} 
\end{quote}
Everything we believe---including what we claim to know---is a belief. QBists  are absolutely right about this. The objective world is what we collectively believe to exist: ``the common external world we have all negotiated with each other.'' The implicit assumption underlying our common external world is not only that the spatiotemporal structure of my perceptual awareness and the logical structure of my thought are the same as the spatiotemporal structure of your perceptual awareness and the logical structure of your thought, that we share the structures to which our basic physical concepts owe their meanings---how else could we be in a position to negotiate a common external world? The implicit assumption is also that my experiences are as definite as yours. I may be ignorant of your experiences, but I cannot doubt the definiteness of your experiences. Wigner may be ignorant of the outcome experienced by his friend, but he should not be cavalier about the definiteness of her experience. By the very nature of our common external world he is required to assign to ``the whole conceptual box'' an incoherent mixture reflecting his ignorance of his friend's actual experience. To treat his own experiences as definite but not those of his friend, to assign to the box a coherent superposition of distinct cognitive states---that would be the solipsism which Wigner feared and sought to avoid by proposing ``that the equations of motion of quantum mechanics cease to be linear, in fact that they are grossly non-linear if conscious beings enter the picture.''

QBists are united in rejecting ``the silly charges of solipsism.'' But if they are to escape these charges, they must do more than acknowledge the fact that ``[m]y experience of you leads me to hypothesize that you are a being very much like myself, with your own private experience''.\cite{MerminQBnotCop} They must also acknowledge that your experiences are as definite as mine. They must draw the dividing line between the classical and the quantum not at the \emph{near} boundary of our common external world, situated between it and each individual user’s subjective experience, nor \emph{within} our common external world, but at its \emph{far} boundary, which separates it from the quantum domain proper, which becomes speakable only via correlations between events that happen or can happen in our common external world. It is our common external world in its entirety that is imbued with the classicality of human sensory awareness.

\section{Objectivation vs. reification}\label{sec.OnotR}
While Bohr failed to realize the full extent of the affinity of his way of thinking with Kant's, QBists fail to realize the full extent of their way of thinking with Bohr's. Thus Fuchs \emph{et al.}\cite{FMS2014}:
\begin{quote}
The Founders of quantum mechanics were already aware that there was a problem. Bohr and Heisenberg dealt with it by emphasizing the inseparability of the phenomena from the instruments we devised to investigate them. Instruments are the Copenhagen surrogate for experience\dots. [They are] objective and independent of the agent using them. 
\end{quote}
QBists, it seems, consistently fail to appreciate the constitutive role played by instruments: instruments define not only the properties that quantum system can possess but also (in conjunction with conservation laws) the kinds of quantum systems that can exist. The reason their outcome-indicating and property-defining attributes are definite is that they are given us in direct sensory experience. While they are objective in the sense that they are experiences that we can objectivize---we can deal with them \emph{as if} they existed independently of being experienced by us---they are not by any means independent of the experiencing and objectivizing agent.%
\footnote{I use the verb ``to objectivize'' in conjunction with ``objectivation,'' leaving the more commonly used verb ``to objectify'' to go with ``objectification,'' which is a stage of the measurement process thought up by $\Psi$-ontologists.}
Again, writes Mermin\cite{MerminQBnotCop}: 
\begin{quote}
Those who reject QBism \dots\ reify the common external world we have all negotiated with each other, purging from the story any reference to the origins of our common world in the private experiences we try to share with each other through language\dots. [B]y ``experience'' I believe [Bohr] meant the objective readings of large classical instruments\dots. Because outcomes of Copenhagen measurements are ``classical,'' they are \emph{ipso facto} real and objective. 
\end{quote}
QBists also seem to consistently overlook the all-important difference between reification and objectivation. For Bohr, as for Kant, the objective world was the common external world we have all negotiated with each other. It contains, \emph{inter alia}, the objective readings of large classical instruments. While outcomes of ``Copenhagen measurements'' are classical because they are (inferred from) objectivized experiences, they are not therefore ``real and objective'' independently of the experiencing and objectivizing subject.%
\footnote{While it is true that those who reify the objective (=objectivized) world will have to reject QBism, it is not necessarily true that those who reject QBism reify our common external world. One can certainly reject some of the (sometimes mutually inconsistent) claims made by QBists without reifying the objective world.}

Bohr was concerned with \emph{objectivation}, the process of representing a shared mental construct as an external reality, which can be dealt with as if it existed independently of the constructing minds. Objectivation means purging any reference to the origins of the ``common body of reality''\cite{FMS2014} constructed by us, not in order to deny its origins in our experiences, but in order to be able to deal with it as common-sense realism does---\emph{as if} it existed independently of our thoughts and perceptions. (That this can no longer be done consistently was the conclusion we reached in Sect.~\ref{sec.amplification}.)

Objectification is not the same as \emph{reification}, which implies complete or willful ignorance of the origins of the objective world in the experiences we try to share with each other through language. Reification is the self-contradictory assertion that the world we perceive exists (just as we perceive it) independently of our perceptions, that the world we mentally construct exists independently of our constructing minds, or that the world we describe exists (just as we describe it) independently of our descriptions.

To Bohr, instruments straddle the far boundary of our common external world. Measurement outcomes are ``classical'' not because they are reified but because they are situated in this intersubjectively constituted world. Instruments make it possible to extend the reach of classical concepts (whose meanings are rooted in the spatiotemporal structure of human sensory perception and the logical structure of human thought) into the non-classical domain via principles of correspondence.%
\footnote{``[Q]uantum mechanics and quantum field theory only refer to individual systems due to the ways in which the quantum models of matter and subatomic interactions are linked by semi-classical models to the classical models of subatomic structure and scattering processes. All these links are based on tacit use of a generalized correspondence principle in Bohr's sense (plus other unifying principles of physics).''  This generalized correspondence principle serves as ``a semantic principle of continuity which guarantees that the predicates for physical properties such as `position', `momentum', `mass', `energy', etc., can also be defined in the domain of quantum mechanics, and that one may interpret them operationally in accordance with classical measurement methods. It provides a great many inter-theoretical relations, by means of which the formal concepts and models of quantum mechanics can be filled with physical meaning''.\cite[pp. XII, 191]{Falkenburg2007}}

\section{Is there a ``world in itself''?}\label{sec.w-i-i}
Kant---the first to make empirical reality the subject of physical science---found it necessary to posit an empirically inaccessible thing or world in itself. This has the power to affect us in such a way that we have the sensations that we do, and that we are able to organize our sensations into objects that interact with each other and change in accordance with physical laws. (It also contains ourselves as transcendental subjects, our free will, and our moral responsibility, but this isn't relevant here.)

Bohr realized that empirical knowledge need not be limited to what is accessible to our senses, and that therefore it does not have to be \emph{solely} a knowledge of interacting, re-identifiable objects and causally connected events. In that he went beyond Kant. But he also realized, with Kant, that what was not directly accessible to our senses could not be expected to conform to the spatiotemporal conditions of human experience, and thus could not be expected to be describable in any language we can understand (apart from the language game of pure mathematics%
\footnote{``To say mathematics is a game is supposed to mean: in proving, we need never appeal to the meaning of the signs, that is to their extra-mathematical application''.\cite{Wittgenstein}}). 
He did not deny the existence of an empirically inaccessible reality; he only denied that physics has anything to do with it.%
\footnote{``We meet here in a new light the old truth that in our description of nature the purpose is not to disclose the real essence of the phenomena but only to track down, so far as it is possible, relations between the manifold aspects of our experience'' [BCW6:\,296]. This has an entirely Kantian ring to it.}

Schr\"odinger has often been depicted as a realist about wave functions. While this was true of the Schr\"odinger of 1926, it does not apply to the Schr\"odinger post 1926, who according to Michel Bitbol\cite{BitbolSPQMp25} adopted a postmodernist stance. This is from a 1950 lecture:
\begin{quote}
we do give a complete description, continuous in space and time without leaving any gaps, conforming to the classical ideal---a description of \emph{something}. But we do not claim that this `something' is the observed or observable facts; and still less do we claim that we thus describe what nature \dots\ really \emph{is}. In fact we use this picture (the so-called wave picture) in full knowledge that it is \emph{neither}.\cite[p.~144, original emphasis]{SchrNGSH}
\end{quote}
Where nature itself was concerned, Schr\"odinger thus agreed with Bohr, who would say (as reported by his assistant Aage Petersen\cite{Petersen1963}) that it is ``wrong to think that the task of physics is to find out how nature is. Physics concerns what we can say about nature.'' What we can say about nature constitutes the empirical reality that for Kant \emph{was} nature.

When it comes to QBism, the situation is less clear. In an early QBist manifesto, Fuchs asked: ``If the quantum state represents subjective information, then how much of its mathematical support structure might be of that same character? Some of it, maybe most of it, but surely not all of it.'' The ``raw distillate'' that is left behind ``when we are finished picking off all the terms (or combinations of terms) that can be interpreted as subjective information \dots\ will be our first glimpse of what quantum mechanics is trying to tell us about \emph{nature itself}''.\cite[emphasis added]{Fuchs_littlemore} Two years later, when talk of ``information'' had been replaced by personalist Bayesian phraseology, Fuchs and Schack wrote that
\begin{quote}
[t]he agent, through the process of quantum measurement stimulates the world external to himself. The world, in return, stimulates a response in the agent that is quantified by a change in his beliefs---i.e., by a change from a prior to a posterior quantum state. Somewhere in the structure of those belief changes lies quantum theory's most direct statement about what we believe of \emph{the world as it is without agents}.\cite[emphasis added]{FS2004} 
\end{quote}
To QBists, quantum mechanics is a generalization of the Bayesian theory of probability. It is a calculus of consistency---a set of criteria for testing coherence between beliefs. In this, the Born rule---formulated in terms of positive-operator-valued measures rather than the standard projection-operator valued ones---is  central. It is seen not merely as a rule for updating probabilities, for getting new ones from old, but as a rule for relating probability assignments and constraining them, a rule that (as they have shown) can be expressed entirely in terms of probabilities: ``The Born Rule is nothing but a kind of Quantum Law of Total Probability! No complex amplitudes, no operators---only probabilities in, and probabilities out''.\cite{Fuchs2010} 

QBists hope to eventually derive the standard Hilbert space formalism from the Born rule, and in doing so puzzle out what is attributable to our way of knowing the world and what is attributable to the world itself. The Born rule has this dichotomic character: it is normative---it guides an agent's behavior in a world that is fundamentally quantum---but it also is an empirical rule. It is a statement about nature itself, indirectly expressed as a calculus of consistency for bets placed on the outcomes of measurements. Writes Fuchs: ``The only piece of the quantum formalism that plays an \emph{objective role} is the normative character of the Born Rule''.\cite[emphasis added]{Fuchs_Notwithstanding}

Mermin,\cite{Mermin2ujm} on the other hand, writes that ``QBists (at least this one) attach no meaning to `the world as it is without agents.' It only means `the common external world we have all negotiated with each other'.'' But then he also writes that ``my understanding of the world rests entirely on the experiences that the world has induced in me throughout the course of my life'',\cite{MerminQBnotCop} and again that ``[t]he world acts on me, inducing the private experiences out of which I build my understanding of my own world''.\cite{Mermin2019} The world that induces private experiences in me---is it our common external world, or is it something like the Kantian thing in itself, which induces the experiences from which we construct our common external world? It seems to me that it has to be the latter, for what induces experiences in us can hardly be the world we construct from our experiences.

In the sections after next I shall present my own take on how our common external world relates to whatever reality lies beyond or at the origin of our common external world.

\section{A reality criterion revisited}\label{sec.realcrit}
Before I encountered QBism, I wrote a series of papers\cite{Mohrhoff-QMexplained,Mohrhoff2009,Mohrhoff2000,Mohrhoff2005,Mohrhoff2011,Mohrhoff2014,Mohrhoff2017} in which I insisted on at least two of the basic tenets of QBism: that quantum mechanics is a probability calculus, and that quantum observables have values only if and when they are actually measured. Probability~1 or the eigenvalue-eigenstate link are not sufficient for ``is'' or ``has.'' Lacking the QBist insight that the irreversibility of measurements and the definiteness of outcomes was attributable \emph{solely} to the definiteness and irreversibility of direct perceptual experience, I proposed a criterion for distinguishing between (i)~observables that have values only when they are measured and (ii)~observables whose values are real \emph{per se} (and thus capable of indicating measurement outcomes). I still regard this criterion as superior to mere appeals to the size or weight of an apparatus, but thanks to QBism I have come to realize that it only allows me to treat the macroworld (herafter defined) as proxy for an intersubjectively constructed world, from which the constructing subjects can detach themselves.

Fuchs\cite{Fuchs_littlemore} once asked an important question, which has already been quoted: ``If the quantum state represents subjective information, then how much of its mathematical support structure might be of that same character?'' Specifically: are the positions on which wave functions depend of the same character? $\Psi$-ontologists are necessarily $xyzt$-ontologists as well, postulating as they must an independently existing spatiotemporal manifold~$\cal M$. This goes as badly as their formulation of the measurement problem, which leads to the disaster of objectification. 

Here is how: Gerhard Hegerfeldt\cite{Hegerfeldt98,Hegerfeldt2001} and David Malament\cite{Malament96} have shown that a free particle, localized at a time $t_1$ in a bounded region~$R_1$, has a non-zero probability to be found at a time $t_2>t_1$ in a bounded region~$R_2$, even if in the time between $t_1$ and $t_2$ no light signal can travel from $R_1$ to~$R_2$. Since this is inconsistent with the theory of relativity, it seems to follow that particles cannot be localized. Having shown that this result also obtains for unsharply localized particles, Hans Halvorson and Rob Clifton\cite{CH2002} concluded that particle talk is ``strictly fictional'': 
\begin{quote}
The argument for localizable particles appears to be very simple: Our experience shows us that objects (particles) occupy finite regions of space. But the reply to this argument is just as simple: These experiences are illusory! Although no object is strictly localized in a bounded region of space, an object can be well-enough localized to give the appearance to us (finite observers) that it is strictly localized.
\end{quote}
What Hegerfeldt, Malament, and Halvorson and Clifton have \emph{actually} shown is that particles are not localizable relative to~$\cal M$. But $\cal M$ is not the expanse in which position measurements are made. Actually measured positions are defined by the sensitive regions of actually existing detectors, and what Halvorson and Clifton have shown for ``objects (particles)'' also holds for such objects as detectors. Neither particles nor detectors are localizable in finite spatial regions of~$\cal M$. Hence what is strictly fictional (i.e., not objectivizable) is the existence of an infinitely or completely differentiated spatiotemporal manifold.

Here, then, is my criterion for distinguishing between observables that only have values if and when they are measured, and observables to which a measure\-ment-independent reality can be attributed. The argument begins by observing that the space that can be objectivized is not intrinsically partitioned. To define a partition of space, we need an array of detectors, and no physically realizable array of detectors can partition space \emph{ad infinitum}, i.e., into infinitesimal regions. The complete differentiation of physical space taken for granted by field theories corresponds to nothing in the objective world.

The next best thing to a sharp trajectory is a trajectory that is so sharp that the bundle of sharp trajectories over which it is statistically distributed is never probed. In other words, the next best thing to an object with a sharp position is a \emph{macroscopic object}, defined as one whose position probability distribution is and remains so narrow that there are no detectors with narrower position probability distributions---detectors that could probe the region over which the object's position statistically extends. 

Macroscopic objects thus follow trajectories that are only counterfactually indefinite. Their positions are ``smeared out'' only in relation to an imaginary spatiotemporal background that is more differentiated than the objectivizable world. The events by which the positions of macroscopic objects are indicated are therefore correlated in ways that are consistent with the laws of motion that quantum mechanics yields in the classical limit. This makes it possible to attribute to the positions of macroscopic objects (collectively referred to as ``the macroworld'') a measurement-independent reality, to regard them as defining the space of positions on which the wave functions of unbound objects%
\footnote{When dealing with an internal relative position (e.g., the position of the electron relative to the proton in the ground state of atomic hydrogen), the positions of the (in this case imaginary) detectors are defined relative to another object (the nucleus), which remains undefined.}
depend, and to use them as apparatus pointers, in the unassailable conviction that they are definite at all times. 

According to Fuchs,\cite{Fuchs2010a} ``there might be uncertainty because the world itself does not yet know what it will give\dots. QBism finds its happiest spot in an unflinching combination of `subjective `probability' with `objective indeterminism'.'' In other words, uncertainty is a consequence of an objective indeterminism. I see it differently. To my mind, indeterminism---lack of predictability not attributable to anyone's ignorance of the facts---is the observable consequence of an objective indefiniteness or indeterminacy (or ``fuzziness'').%
\footnote{\label{note.stability}It is this indefiniteness, made irreducible by the uncertainty relations, that is at least in part responsible for the existence of stable atomic ground states, and ultimately for the stability of matter and thus the existence of matter as we know it.}
But whereas the objective indefiniteness of a quantum observable is revealed by the unpredictability of the position of the ``pointer'' upon measurement, the indefiniteness of the (position of the) pointer is never revealed. That is: it cannot be objectivized; it is not an objective indefiniteness; it corresponds to nothing in the objective world.

\section{The mystery of identity}\label{sec.identity}
{\leftskip\parindent{No acceptable explanation for the miraculous identity of particles of the same type has ever been put forward. That identity must be regarded, not as a triviality, but as a central mystery of physics.\par\hfill--- Charles W. Misner, Kip S. Thorne, and John A. Wheeler}\cite{Misneretal1215}\par}
\medskip\noindent
My approach to extracting essential information about what lies beyond the objective (=objectivizable) world from correlations between measurement outcomes relies on Feynman's formulation of the theory.\cite{FHS} This is based on summations over alternatives, which are defined as sequences of outcomes of (performed or unperformed) measurements. Suppose, for example, that we%
\footnote{The plural is justified by the fact that measurements take place in our common external world.}
perform a series of position measurements, and that each measurement yields exactly one outcome, i.e., each time exactly one detector clicks. This would be evidence of a conservation law and could be construed as evidence of a persistent quantum object---not an object that \emph{causes} clicks (Ulfbeck and Bohr are absolutely right about this) but an object whose successive positions are indicated by the clicks. We would then be able to construe the clicks as measurements of the position of a quantum object and to think of the clicking devices as detectors for such objects.

If each time exactly $n$ detectors click, we have evidence that the number of simultaneous detector clicks is a conserved quantity, but this cannot be construed as evidence of a fixed number of re-identifiable quantum objects unless further conservation laws are in force. This is where Feynman's pair of rules of summation becomes important. If the alternatives are indistinguishable, we assign probabilities by adding their amplitudes and calculating the absolute square of the result. If the alternatives are distinguishable, we assign probabilities by calculating the absolute square of each amplitude and adding the results. In the second case the distinctions we make between the alternatives can be objectivized, for example because the simultaneous clicks are of different types, individuating different Lockean substances (such as electrons and protons), and because there is a conservation law for each of these Lockean substances.%
\footnote{Here the reader may want to revisit the discussion in Sect.~\ref{sec.QsysQpha}. A single click does not usually announce the type to which it belongs---whether it's an electron click or a proton click. In general the type of a click has to be inferred from a sequence of clicks. A sequence of clicks makes it possible to determine such quantities as the radius of curvature of a particle's track in a magnetic field, a particle's time of flight, a particle's kinetic energy, or a particle's energy loss through ionization and excitation. Measuring three of these four quantities is sufficient in principle to positively identify the particle type, which then makes it possible to classify the individual clicks.\cite{GS2008}}
In this case the behavior of the detectors can be construed as indicating the successive positions of $n$ re-identifiable quantum objects. 

In the first case, nothing in our common external world indicates which quantum object is which. In other words, the distinctions we make between the alternatives cannot be objectivized. We are then in the presence of a \emph{single} quantum object, which is instantiated but not individuated by the clicks. If we nevertheless think of the clicks as indicating the positions of quantum objects, we must think of the objects instantiated by the clicks as identical, and this not in the weak sense of exact similarity but in the strong sense of \emph{numerical identity}. They are the same object in $n$ different places. What is signaled by the detectors that click is the presence of one and the same object in each of their respective sensitive regions.

But why should we treat a positions differently from other properties, such as the properties that make electrons distinguishable from protons? Is there any compelling reason to believe that the numerical identity of quantum objects in different places ceases when it ceases to have observable consequences owing to the presence of ``identity tags''? I can think of no such reason. I am therefore prepared to defend the following claim: a quantum object observed here with these properties and a quantum object observed there with those properties are one and the same thing. It appears to us here with these properties and there with those properties.

Kant did not stop at saying that if I see a desk, there is a thing in itself that has the power to appear as a desk, and if I see a chair in front of the desk, there is another thing in itself that has the power to appear as a chair. For him, there was only {one} {thing in itself}, which affects us in such a way that we see both a desk and a chair in front of the desk. What would be news to him is that all is not desks and chairs. In addition to phenomena in the traditional sense there are quantum phenomena. In addition to the universal context of human experience, and within the same, there are experimental contexts instantiating Lockean substances. If we insist on thinking of these instantiations as things, then quantum mechanics strongly suggests that the thing we observe here with these properties and the thing we observe there with those properties is what Kant would have called the thing in itself. It appears to us here as an electron and there as a proton.

What we have learned from Kant is this: if our minds are to be able to ``work up the raw material of sensible impressions into a cognition of objects'' [CPR\,136], the system of objects he called ``nature'' must obey certain laws. What Kant could not tell us is how the thing in itself affects us in such a way that we are able to work up the raw material of sensible impressions into a cognition of objects. This is where quantum mechanics comes in. The knowledge it provides does not concern laws that a world of experience objects obeys. It concerns how something corresponding to Kant's thing in itself causes itself to be experienced as a world of objects. It touches on the age-old subject of how a One becomes Many, but it does not concern the coming into being of a world that exists independently of experiencing subjects, agents, or users, and it does not concern how such a world comes to be reflected in our minds. It concerns the coming into being of an \emph{experienced} world, without interposition of an unexperienced world.

\section{The poises of creative awareness}\label{sec.poises}
Schr\"odinger held that the ``extensive agreement or parallelism'' between our hermetically separated ``spheres of consciousness'' can only be explained by either of two ``irrational, mystical hypotheses.'' In Sect.~\ref{sec.3realisms} we saw what he thought of one of them. The other, which he endorsed, was that ``we are all really only various aspects of the One''\cite{SchrWhatIsReal}: the multiplicity of minds, he wrote,\cite{SchrLifeMindMatterAPOM}  ``is only apparent, in truth there is only one mind. This is the doctrine of the Upanishads. And not only of the Upanishads''.(The Upanishads are ancient Sanskrit texts which contain many of the central concepts and ideas of Indian philosophy.) The One here refers to the Ultimate Subject, from which we are separated by a veil of self-oblivion. The same veil (according to the Upanishads) also prevents us from perceiving the Ultimate Object, as well as its identity with the ultimate subject.%
\footnote{Schr\"odinger\cite{SchrLifeMindMatterPO} adds that if ``to Western thought this doctrine has little appeal,'' it is because our science ``is based on objectivation, whereby it has cut itself off from an adequate understanding of the Subject of Cognizance, of the mind,'' and that ``this is precisely the point where our present way of thinking does need to be amended, perhaps by a bit of blood-transfusion from Eastern thought. That will not be easy, we must beware of blunders---blood-transfusion always needs great precaution to prevent clotting. We do not wish to lose the logical precision that our scientific thought has reached, and that is unparalleled anywhere at any epoch.''}

If at bottom we are all the {same} subject (without being aware of it, except by a genuinely mystical experience that is hard to come by), we can conceive of two poises of consciousness or two modes of experience, one in which the One manifests the world to itself perspectivally, as if experienced by a multitude of subjects from a multitude of locations within the world, and one in which the One manifests the world to itself aperspectivally, as if experienced from no particular location or from everywhere at once.%
\footnote{Such an aperspectival experience features prominently in the work of Jean Gebser\cite{GebserEPO,Mohrhoff-Gebser} and in the philosophy of Sri Aurobindo.\cite{SA-TLD} Here is an account of an experience of this kind: ``It is as if the consciousness was not in the same position with regard to things---I do not know how to say it\dots. The ordinary human consciousness, even when it has the widest ideas, is always at the centre, and things are like this (\emph{gesture of convergence from all sides})\dots. I believe this is how it is best expressed: in the ordinary human consciousness one is at a point and all things exist in their relation to this point of consciousness. And now, the point exists no more\dots. So, my consciousness is in the things---it is not something which is receiving''.\cite{NOTW277} In other words, the subject is where its objects are; it lacks the distantiating viewpoint of our perspectival outlook.}
And if we distinguish between two such poises, we can, so Schr\"odinger affirms, understand the origin of the agreement between our private external worlds. His assertion that the agreement ``between the content of any one sphere of consciousness and any other'' was established by language (and corresponding declarations by QBists) leaves unexplained the agreement between my \emph{sense impressions} and yours, without which it would be impossible for my \emph{description} of my impressions to agree with your \emph{description} of your impressions. The reason why my {sense impressions} agree with yours (to the extent that they do) is that we do, in fact, experience the same world. We experience perspectively the world that the One manifests to itself aperspectively. 

If there is no unexperienced world, the question how such a world can come to be experienced or reflected in a conscious mind does not arise. In Schr\"odinger's own words,
\begin{quote}
to say \dots\ that the becoming of the world is reflected in a conscious mind is but a clich\'{e}, a phrase, a metaphor that has become familiar to us. The world is given but once. Nothing is reflected. The original and the mirror-image are identical. The world extended in space and time is but our representation. Experience does not give us the slightest clue of its being anything besides that.\cite{SchrLifeMindMatterAPOM}
\end{quote}
The question that arises instead concerns the relation between the two poises of consciousness or modes of experience. According to the Upanishads, all knowledge (or experience, or awareness) is based on identity. The One is indistinguishably both a consciousness that contains objects and a substance that constitutes objects. But if the One adopts a multitude of localized standpoints within the world it manifests to itself, knowledge by identity takes the form of a direct knowledge: each individual knows the others directly, without mediating representations.%
\footnote{In the original, aperspectival poise of relation between the One and the world, the One is coextensive with the world. As yet no distances exist between the knower and the known. In other words, space as we know it does not yet exist. The familiar dimensions of phenomenal space (viewer-centered depth and lateral extent) come into being in this secondary poise, in which the One views the world in perspective. Objects are then seen from ``outside,'' as presenting their surfaces. Concurrently the dichotomy between subject and object becomes a reality, for a subject identified with an \emph{individual} form cannot be overtly identical with the substance that constitutes \emph{all} forms.}
And if the localized subject identifies itself with an individual \emph{to the exclusion of all other individuals}, direct knowledge of objects takes the form of an indirect knowledge---a knowledge mediated by representations. It becomes a direct knowledge of some of the individual's attributes---think electrochemical pulses in brains---that gets transformed into a knowledge of ``external'' objects with the help of a subliminal direct knowledge.

\section{Indirect knowledge}\label{sec.indirect}
While it used to be said that qualities are ``nothing but'' quantities (e.g., colors are ``nothing but'' frequencies or reflectances), it may be much closer to the truth to say that quantities are nothing but means of manifesting qualities. What is not sufficiently appreciated is that not only the sensations of color, sound, taste, smell, and touch fail to be reducible to quantities. Our very experience of space and time is qualitative and therefore equally irreducible to quantities. 

Like the color of a Burmese ruby, spatial extension is a quality that can only be defined by ostentation---by drawing attention to something of which we are directly aware. If you are not convinced, try to explain to my friend Andy, who lives in a spaceless world, what space is like. Andy is good at math, so he understands you perfectly if you tell him that space is like a set of all triplets of real numbers. But if you believe that this gives him a sense of the expanse we call space, you are deluding yourself. \emph{We} can imagine triplets of real numbers as points embedded in space; he cannot. \emph{We} can interpret the difference between two numbers as the distance between two points; he cannot. At any rate, he cannot associate with the word ``distance'' the phenomenal remoteness it conveys to us.%
\footnote{The same point was made by Hermann Weyl\cite{Weyl1922} when he wrote that geometry ``contains no trace of that which makes the space of intuition what it is in virtue of its own entirely distinctive qualities which are not shared by `states of addition-machines' and `gas-mixtures' and `systems of solutions of linear equations'.''}

Much the same goes for time. Time passes, and the only way to know this is to be aware of it. This is what St. Augustine meant when he wrote, ``What, then, is time? If no one asks me, I know; if I wish to explain to him who asks, I know not.'' That the passingness of time is another quality which cannot be defined in quantitative or mathematical terms, is obvious from the fact that we cannot measure the speed at which it passes. (One second per second?)

Neuroscience has learned a great deal about how the brain extracts information from images falling on our retinas.\cite{Hubel95,Enns04,Koch2004} This information is encoded in patterns of electrochemical pulses, and these patterns need to be interpreted in order to be experienced as (or give rise to experiences of) objects in an external world extended in space and time. The decoding or interpretation of these firing patters presupposes acquaintance with the expanse of space and the passingness of time, and such acquaintance is not something that neural processes can provide. So the question is not only ``whence the sensory qualities?'' but also ``whence our forms of perception?''

There are sound reasons to doubt that the empirically accessible correlations between measurable brain function and qualitative experience will ever fall within the purview of rational scientific explanation.\cite{McGinn99,Nagel_Nowhere} After all, as Maurice Merleau-Ponty\cite{MerleauPonty1983} and Karl Jaspers\cite{Jaspers1947} have pointed out, the existence of correlations between sensory experiences and neural processes is itself a fact of sensory experience. The empirically known correlations exist between experiences and therefore cannot be invoked to explain how neural processes give rise to experiences. Sensory experiences do not give rise to sensory experiences.

To clearly get this, imagine a neuroscientist, Alice, who observes a specific complex of neural processes in Bob's visual cortex whenever she sees that a green apple is located in Bob's visual field. Something in her experience of Bob's brain correlates with something in her experience of what Bob is looking at. If Bob tells her that he, too, perceives a green apple, it confirms the existence of a green apple in a shared objective world. What it does not confirm is the existence of a real apple that causes both Alice and Bob to perceive an apple, nor the belief that Bob's brain---as experienced and studied by Alice---serves as a link in a causal chain that connects a real apple in a mind-independent external world to Bob's experience of an apple. 

Again, when we say ``this is a green apple,'' we do not state the correspondence of a perception to a thing-in-itself. While our judgment that this is a green apple goes beyond what is \emph{immediately} given to us, it does not reach beyond what is given to us. It merely involves the claim that this thing is of much the same color, shape, and consistency as the things we previously judged to be green apples, or the claim that this particular {experience} is of the same kind as experiences we previously referred to as ``green apples.'' It involves the correspondence between ``green apple experienced here and now'' and ``green apple experienced there and then.'' Representations are \emph{re}-presentations of experiential material that was present at some other time. They are objective in the sense of being recognizable invariants of experience. 

\begin{figure}[t]\begin{center}
\includegraphics[width=3in]{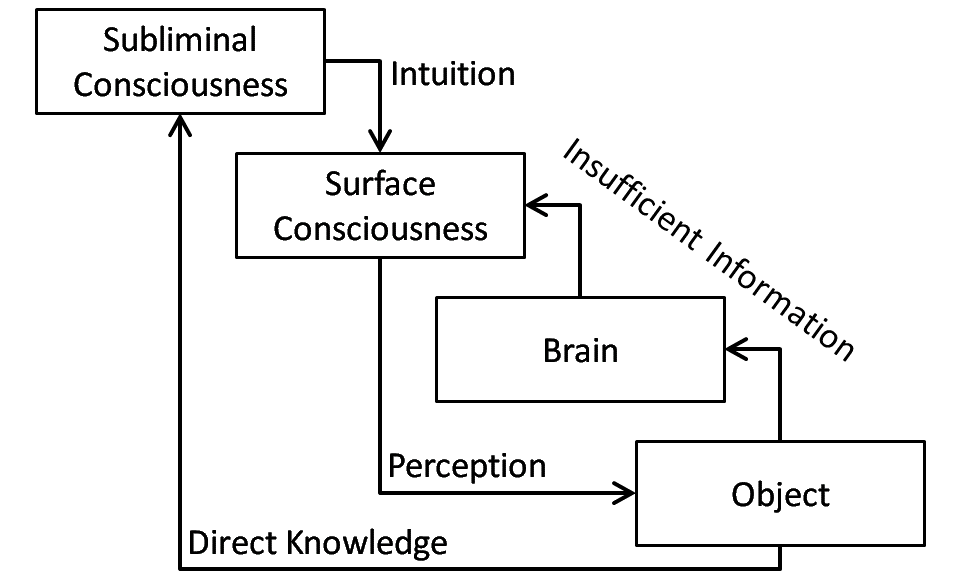}
\caption{Indirect knowledge: a flowchart}\label{fig.subliminal}
\end{center}\end{figure}

On the other hand, if the ``irrational, mystical'' hypothesis we are here exploring is correct, then the incomplete information provided by neural firing patterns is supplemented by a subliminal direct knowledge, and in this case Bob's experience of an apple is veridical,%
\footnote{To be precise: as veridical as a knowledge mediated by representations can be.}
and Alice's experience of neural firing patterns in Bob's brain is a veridical experience of representations mediating Bob's experience of an apple. Indirect knowledge is the meeting point of information flowing inward from the object and information flowing outward from a subliminal self (Fig.~\ref{fig.subliminal}).

In the philosophy of mind, the problem of intentionality is as formidable as the problem of qualia. ``Intentionality'' denotes the fact that, instead of perceiving internal representations, we perceive external objects. Both problems resolve themselves if we accept the fundamental affirmations of the Upanishadic theory of existence. Whatever is missing from the internal representations---intentionality, qualia, including our forms of perception---is supplied by a subliminal subject's direct awareness of objects. This direct awareness is rooted in the Ultimate Subject's identity with the Ultimate Object, which appears to us here with these properties and there with those properties. In the words of Sri Aurobindo (arguably the most competent modern interpreter of Upanishadic thought):
\begin{quote}
In the surface consciousness knowledge represents itself as a truth seen from outside, thrown on us from the object, or as a response to its touch on the sense, a perceptive reproduction of its objective actuality\dots. Since it is unable to \dots\ observe the process of the knowledge coming from within, it has no choice but to accept what it does see, the external object, as the cause of its knowledge\dots. In fact, it is a hidden deeper response to the contact, a response coming from within that throws up from there an inner knowledge of the object, the object being itself part of our larger self. \cite[pp. 560--61]{SA-TLD}
\end{quote}

\section{Why quantum mechanics?}\label{sec.whyqm}
{\leftskip\parindent{It seems clear that quantum mechanics is fundamentally about atoms and electrons, quarks and strings, not those particular macroscopic regularities associated with what we call \emph{measurements} of the properties of these things. But if these entities are not somehow identified with the wave function itself---and if talk of them is not merely shorthand for elaborate statements about measurements---then where are they to be found in the quantum description?\par\smallskip\hfill--- Sheldon Goldstein}\cite{Goldstein2017}\par}
\medskip\noindent
Here is what I believe to be the most important message that quantum mechanics has for us: the theory does not primarily concern the world that the One manifests to itself. It primarily concerns \emph{how} the One manifests the world to itself---and therefore to us, inasmuch as we ``are all really only various aspects of the One.'' Rather than being constituent parts of the manifested world, subatomic particles, atoms, and (to some extent) molecules are instrumental in the manifestation of the world (to us). 

We keep looking for the origin of the universe at the beginning of time, but this is an error of perspective. The origin of the manifested world is the One, transcendent of spatial and temporal distinctions, and the manifestation of the world is an (atemporal) series of transitions away from the unity of the One qua Ultimate Object to the multiplicity of a world that allows itself to be described (not in terms of classical physics but) in the object-oriented language of classical physics. 

The first transition takes us from the One (who of course is formless) to an apparent multitude of formless ``ones.'' By entering into reflexive spatial relations, the Ultimate Object gives rise to (i)~what looks like a multitude of relata if the reflexive quality of the relations is ignored, and (ii)~what looks like a substantial expanse if the spatial quality of the relations is reified. The relata are usually referred to as ``fundamental particles'' and regarded as the ``ultimate constituents of matter''.%
\footnote{Fundamental particles are often characterized as pointlike, and many  physicists take this to be the literal truth. What it actually means is that a fundamental particle lacks internal relations. Lacking internal relations, it also lacks a form, inasmuch as forms resolve themselves into sets of internal spatial relations. And if we conceive of objective space as the totality of objective or objectivizable spatial relations, then a fundamental particle, lacking internal relations, cannot even be said to exist or be contained in space. (A reminder: in Sect.~\ref{sec.realcrit} we arrived at the conclusion that the existence of an infinitely and completely differentiated spatiotemporal manifold must be regarded as an illusion.) There is even a sense in which space is internal to each fundamental particle, inasmuch as each fundamental particle is the One qua Ultimate object, and each spatial relation is a relation between the One and the One.}
The result of this transition is probed by high-energy physics and known to us through correlations between detector clicks (i.e., in terms of transition probabilities between in-states and out-states). 

Forms emerge as sets of more or less indefinite spatial relations between formless and numerically identical relata, i.e., between the One and the One. The forms of nucleons, nuclei, and atoms ``exist'' in probability spaces of increasingly higher dimensions. At energies low enough for atoms to be stable, we are dealing with Lockean substances that can be described in terms of correlations between the possible outcomes of unperformed measurements. (Recall note~\ref{note.stability}.) 

At the penultimate stage of the (atemporal) process of manifestation there emerges a kind of form that can be visualized, and this not merely as a distribution over some probability space. What makes the atomic configurations of molecules visualizable is that the indefiniteness of the distance $d$ between any pair of bonded atoms, as measured by the standard deviation of the corresponding probability distribution, is significantly smaller than the mean value of~$d$.

If the manifestation of the world consists in a progressive transition from the undifferentiated unity of the One to a multitude of distinguishable objects with definite properties, via formless particles, non-visualizable atoms, and partly visualizable molecules, the question arises as to how the intermediate stages of this transition are to be described---the stages at which distinguishability and definiteness are incompletely realized. The answer is that whatever is not completely distinguishable or definite can only be described in terms of probability distributions over what \emph{is} completely distinguishable and definite---i.e., over the possible outcomes of measurements. What is instrumental in the manifestation of the world can only be described in terms of correlations between events that happen (or could happen) in the manifested world. This, I believe, is why the general theoretical framework of contemporary physics is a probability calculus, and why the events to which it serves to assign probabilities are measurement outcomes.

\section{The adventure of evolution}\label{sec.evolution}
But why should the ultimate subject not only adopt a multitude of standpoints but also identify itself with each to the apparent exclusion of the others? The answer hinges on the evolutionary character of this world. From the point of view of the Upanishads, evolution presupposes involution. By a multiple concentration of consciousness the ultimate subject assumes a multitude of vantage points, and by a further \emph{exclusive} concentration of consciousness the individual subject loses sight of its identity with the other subjects and, as a consequence, loses access to the aperspectival view of things. The direct self-knowledge of the One thereby becomes implicit, or involved, in an indirect or representative knowledge. 

While our state of being and knowing in some respects corresponds to this poise of relation between self and world, we are evolved from a condition of maximal involution rather than ``devolved'' by a multiple exclusive concentration of a consciousness that is one with existence. Because of the identity of the Ultimate Subject with the Ultimate Object, the consciousness by which the former creates its content is also the force by which the latter creates forms. In the original, unitary and aperspectival poise of awareness, force is rather implicit in consciousness. At the other end of the spectrum of involution, consciousness becomes implicit (or involved) in the force that creates forms. But this is not all. As there is a world-transcending state of self-awareness,%
\footnote{``For at the gates of the Transcendent stands that mere and perfect Spirit described in the Upanishads, luminous, pure, sustaining the world but inactive in it, \dots the transcendent Silence. And the mind when it passes those gates suddenly, without intermediate transitions, receives a sense of the unreality of the world and the sole reality of the Silence which is one of the most powerful and convincing experiences of which the human mind is capable. Here, in the perception of this pure Self or of the Non-Being behind it, we have the starting-point for a second negation,---parallel at the other pole to the materialistic, but more complete, more final, more perilous in its effects on the individuals or collectivities that hear its potent call to the wilderness,---the refusal of the ascetic. It is this revolt of Spirit against Matter that for two thousand years \dots has dominated increasingly the Indian mind\dots. And through many centuries a great army of shining witnesses, saints and teachers, names sacred to Indian memory and dominant in Indian imagination, have borne always the same witness and swelled always the same lofty and distant appeal,---renunciation the sole path of knowledge, acceptation of physical life the act of the ignorant, cessation from birth the right use of human birth, the call of the Spirit, the recoil from Matter. For an age out of sympathy with the ascetic spirit \dots\ it is easy to attribute this great trend to the failing of vital energy in an ancient race\dots. But we have seen that it corresponds to a truth of existence, a state of conscious realisation which stands at the very summit of our possibility.'' \cite[pp. 26--27]{SA-TLD}}
so there is state of inconscience transcendent of both self and world. If involution begins with a multiple localization of the Ultimate Subject resulting in multitude of localized subjects, it ends not with the involution of consciousness in force, nor even with the involution of force in form, but in the involution of consciousness and formative force in a multitude of formless entities. And since formless entities are indistinguishable and therefore---by the principle of the identity of indiscernibles---numerically identical, evolution ends with the Ultimate Object apparently deprived of its inherent consciousness and self-determining force. This (or something very much like it) is how the stage for the drama of evolution was set.

But what can justify this adventure, considering all the pain and suffering that (in hindsight) it entails? Certainly not an extra-cosmic Creator imposing these evils on his creatures. But the One of the Upanishads is no such monster; it imposes these things on itself. But still---why? Here goes:
\begin{quote}
a play of self-concealing and self-finding is one of the most strenuous joys that conscious being can give to itself, a play of extreme attractiveness. There is no greater pleasure for man himself than a victory which is in its very principle a conquest over difficulties, a victory in knowledge, a victory in power, a victory in creation over the impossibilities of creation\dots. There is an attraction in ignorance itself because it provides us with the joy of discovery, the surprise of new and unforeseen creation\dots. If delight of existence be the secret of creation, this too is one delight of existence; it can be regarded as the reason or at least one reason of this apparently paradoxical and contrary Lila. \cite[pp.\ 426--27]{SA-TLD}
\end{quote}
\emph{L\={\i}l\={a}} is a term of Indian philosophy which describes the manifested world as the field for a joyful sporting game made possible by self-imposed limitations. ``Conscious being'' is Sri Aurobindo's term for the ultimate subject or consciousness that is one with the ultimate object or being. ``Delight of existence'' is the third of the three terms by which the One is described in the Upanishads. If as being (\emph{sat}) the One constitutes the world and as consciousness (\emph{chit}) it contains it, as an infinite Quality, Value, and Delight (\emph{\={a}nanda}) it experiences and expresses itself in it. 

In a materialistic framework of thought, what ultimately exists is a multitude of entities (fundamental particles, spacetime points, whatever) lacking intrinsic quality or value. In many traditions this multiplicity is fittingly referred to as ``dust.'' Such a framework leaves no room for a non-reductive account of quality and value. In an Upanishadic framework, quality and value and their subjective counterpart joy or delight are at the very heart of reality. In such a framework accounting for the qualitative content of consciousness poses no difficulty, since it is simply the finite expression or manifestation of the infinite quality inherent in reality itself.

\section{A possible future}\label{sec.future}
Because what has been involved is an unlimited consciousness and force, evolution is far from finished. What has yet to evolve is a consciousness that is not exclusively concentrated in each individual, a consciousness aware of the mutual identity of all individuals, a consciousness no longer confined to the perspectival outlook of a localized being but capable of integrating its perspectival viewpoint with the aperspectival outlook of the Ultimate Subject.

In recent years, many philosophers have put a high priority on providing a reductionist account of intentional categories such as beliefs and desires. There is an unmistakable tone of urgency in the relevant literature. Apparently something dreadful will follow if the program of ``naturalizing'' the intentional does not succeed---something more dreadful (to the physicalist) than what would follow if the program were to succeed.\cite{StichLaurence} If the program were to succeed, it would not be true that my wanting something is causally responsible for my reaching for it, that my itching is causally responsible for my scratching, and that my believing something is causally responsible for my saying it. And ``if none of that is literally true,'' Jerry Fodor concludes,\cite{Fodor90} ``then practically everything I believe about anything is false and it's the end of the world''.%
\footnote{ In other words,\cite{Fodor87}, ``if commonsense intentional psychology really were to collapse, that would be, beyond comparison, the greatest intellectual catastrophe in the history of our species; if we're that wrong about the mind, then that's the wrongest we've ever been about anything. The collapse of the supernatural, for example, didn't compare; theism never came close to being as intimately involved in our thought and our practice---especially our practice---as belief/desire explanation is. Nothing except, perhaps, our commonsense physics---our intuitive commitment to a world of observer-independent, middle-sized objects---comes as near our cognitive core as intentional explanation does.''}

What, then, shall we make of the neuroscientific and psychological evidence that, along with philosophic analysis and phenomenological introspection, supports the conclusion that the folk model of free will is seriously flawed?%
\footnote{Because classical physics was taken by many to imply determinism, it is not surprising that the indeterminism of quantum physics has been invoked as the physical basis of free will. One of the first to suggest that quantum indeterminism makes room for mental causation---the determination of physical events by irreducibly mental ones---was Pascual Jordan.\cite{Jordan1944} And it was none other than Schr\"odinger\cite[pp.\ 164--66]{SchrNGSH} who pointed out the flaws in Jordan's reasoning, considering it ``to be both physically and morally an impossible solution.''}
What the opponents as well as the proponents of the libertarian (non-compa\-ti\-bilist) version of free will miss, is the subliminal self and the role it plays in both volition and cognition. Let us assume, for the sake of the argument, that Sri Aurobindo is right when he writes that ``to establish an infinite freedom in a world which presents itself as a group of mechanical necessities \dots\ is offered to us as \dots\ the goal of Nature in her terrestrial evolution''.\cite[p.~4]{SA-TLD} There is only one way in which infinite freedom can be attained, and that is to become one with the ultimate determinant of this evolving manifestation. We are in possession of genuine freedom to the extent that we are consciously and dynamically identified with this ultimate determinant. Absent this identification, our sense of being the proud owner of a libertarian free will is, not indeed a complete illusion, but a misappropriation of a power that belongs to our subliminal self, a power that more often than not works towards goals at variance with our conscious pursuits.

Just as the One adopts several cognitive poises of relation between subject and object, so it adopts several dynamic poises. At the extreme point of involution, we have an apparent multitude of formless particles governed by laws that serve to set the stage for the adventure of evolution.%
\footnote{Setting the stage for evolution arguably requires objects that are spatially extended (they ``occupy'' finite volumes) and are sufficiently stable (they neither explode nor collapse as soon as they are formed). And because the stage was set by an involution of consciousness and formative forms in an apparent multitude of formless entities, these objects appear to be ``made of'' finite numbers of particles that do not ``occupy'' space. As I have argued previously,\cite{Mohrhoff-QMexplained,Mohrhoff-JustSo,Mohrhoff-book25} the existence of such objects not only implies the validity of quantum mechanics but also goes a long way toward establishing the other well-tested laws of physics (the standard model and general relativity).}
That's all they do. They do not direct what happens on the stage. Evolution does not happen without active modifications (or, God forbid, ``violations'') of the laws that served to set the stage. If (as we presently assume) the adventure was made possible by an unlimited force subjecting itself to these laws, its ability to modify them should be a no-brainer. 

So why do we not have irrefutable evidence that such modifications occur? For one, because we tend to look for them where they do not occur (e.g., at the origin of our supposedly free choices). For another, because of the Houdiniesque nature of this manifestation. ``If delight of existence be the secret of creation,'' and if the joys of winning victories, conquering difficulties, overcoming obstacles, and discovering the unknown be possible expressions of this delight, then there have to be serious limitations, initially and for a long time, on the power to modify the laws, and this makes it virtually impossible for us (given the means at our disposal) to discern where and when such modifications occur.%
\footnote{What about the alleged causal closure of the physical? Since the doctrine of causal closure is tantamount to the claim that modifications of the laws of physics cannot occur, it obviously cannot be invoked as an \emph{argument} against the occurrence of such modifications.}

Each of the major evolutionary transitions entails a displacement of the boundary between what is overtly at play and what acts subliminally. To see what I mean, think of the creative process by which the infinite Quality at the heart of reality expresses itself in finite forms as encompassing two intermediate stages:

\medskip\centerline{\emph{Infinite Quality $\rightarrow$ Expressive Idea $\rightarrow$ Formative Force $\rightarrow$ Finite Form}}

\medskip\noindent The boundary can be located between any two stages. When life appears, what essentially begins to be overtly at play is the formative force whose principal purpose is to execute expressive ideas, while the power to form expressive ideas acts subliminally if at all. When mind appears, what essentially comes openly into play is the power whose principal purpose is to develop Quality into expressive ideas, while the Quality to be expressed continues, for the most part, to dwell in the subliminal recesses of existence. Nor can the formative force, when it appears, begin at once to execute expressive ideas; it first has to fashion the necessary anatomical and physiological instrumentation. Nor can the power to form expressive ideas act without the requisite instrumentation, and when it begins to act, it must attend to the needs of self-preservation and self-development before it can turn to loftier pursuits.

What will happen if the infinite Quality at the heart of reality finally comes in front and the entire creative process becomes conscious and deliberate? Will there remain any need for complex anatomies and physiologies? To put it bluntly, will the complete embodiment of the original Consciousness and Force dispense with muscles, bones, stomachs, hearts, and lungs? Will the world be experienced by its future inhabitants without the need for mediating representations and thus without the need for a brain? When the creative process is no longer broken into an automatic or mechanical action on the surface and a subliminal modifying action, when the previously automatic action is fully integrated into the formerly modifying action, when the limited action and the limiting action are fused into one unlimited action, the anatomical and physiological instrumentation will have served its purpose and can be dispensed with, for its evolution was necessitated by the limitations that the One had imposed on itself for the purpose of instituting the adventure of evolution, and these limitations no longer exist.

If any or all of this sounds preposterous, it is in large part because our theoretical dealings with the world are conditioned by the manner in which we, at this particular evolutionary juncture, experience the world. We conceive of the evolution of consciousness, if not as a sudden lighting up of the bulb of sentience, then as a progressive emergence of ways of experiencing a world that exists independently of how it is experienced, but which nevertheless is structured or constituted more or less as it is experienced by us. In reality there is no world that exists independently of how it is experienced. There are only different ways in which the One manifests itself to itself. 

The different ways in which the One has so far manifested itself to itself have been painstakingly documented by Gebser.\cite{GebserEPO,Mohrhoff-Gebser} One way to characterize the structures of human consciousness that have emerged or are on the verge of emerging, is in terms of their dimensionality. An increase in the dimensionality of the consciousness to which the world is manifested is tantamount to an increase in the dimensionality of the manifested world. 

Consider, by way of example, the consciousness structure that immediately preceded the still dominant one. One of its characteristics was the notion that the world is enclosed in a sphere, with the fixed stars attached to its boundary, the firmament. \emph{We} cannot but ask: what is beyond that sphere? Those who held this notion could not, because for them the third dimension of space---viewer-centered depth---did not at all have the reality it has for us. Lacking our sense of this dimension, the world experienced by them was in an important sense two-dimensional. This is why they could not handle perspective in drawing and painting, and why they were unable to arrive at the subject-free ``view from nowhere'',\cite{Nagel_Nowhere} which is a prerequisite of modern science. All this became possible with the consolidation, during the Renaissance, of our characteristically three-dimensional structure of consciousness.

Our very concepts of space, time, and matter are bound up with, are creations of our present consciousness structure. It made it possible to integrate the location-bound outlook of a characteristically two-dimensional consciousness into an effectively subject-free world of three-dimensional objects. Matter as we know it was the result.%
\footnote{As was the so-called mind-body problem.}
It is not matter that has created consciousness; it is consciousness that has created matter, first by carrying its multiple exclusive concentration to the point of being involved in an apparent multitude of formless particles, and again by evolving our present mode of experiencing the world. Ahead of us lies the evolution of a consciousness structure---and thus of a world---that transcends our time- and space-bound perspective. Just as the mythological thinking of the previous consciousness structure could not foresee the technological explosion made possible by science, so scientific thinking cannot foresee the consequences of the birth of a new world, brought about, not by technological means, but by a further increase in the dimensionality of the evolving consciousness.

In this new world, will consciousness still depend on a brain? Will bodies still depend on organs and cells? We can envision a new perception in which our anatomical insides are replaced by the previously hidden truths of our individual existences, a new seeing in which our true natures---our mutually complementary embodiments of the One---are revealed. We find it harder to accept that the new way of seeing will be the new way of being, and that in this new way of seeing and being our anatomical insides will have disappeared, discarded like a scaffolding or the chrysalis of a butterfly.

\end{document}